\documentclass[10.75pt, a4paper]{article}
\usepackage{titlesec}
\titleformat{\section}
  {\bf\sffamily}
  {\thesection. }
  {5pt}
  {\MakeUppercase}
\renewcommand{\thesection}{\Roman{section}} 

\titleformat{\subsection}
  {\bf\sffamily}
  {\thesubsection. }
  {5pt}{}
  
\renewcommand{\thesubsection}{\Alph{subsection}}

\usepackage[affil-it]{authblk} 
\usepackage{etoolbox}
\usepackage{lmodern}

\usepackage{eurosym}

\usepackage[utf8]{inputenc}
\usepackage{geometry}
\usepackage[hidelinks]{hyperref}
\usepackage[citestyle=numeric,style=phys,biblabel=brackets,backend=bibtex,sorting=none,url=false,doi=false, natbib]{biblatex}
\bibliography{bibliography}
\DefineBibliographyStrings{english}{andothers={\itshape et\addabbrvspace al\adddot}}

\usepackage{csquotes}
\usepackage[]{authblk} 
\usepackage{etoolbox}
\usepackage{lmodern}

\usepackage{amsmath}
\usepackage{enumerate}
\usepackage{enumitem}
\usepackage{graphicx}
\usepackage{siunitx}
\usepackage{float}
\usepackage[]{parskip} 

\makeatletter
\patchcmd{\@maketitle}{\LARGE \@title}{\fontsize{16}{19.2}\selectfont\@title}{}{}
\makeatother

\usepackage[normalem]{ulem}

\usepackage{graphicx}
\usepackage{dcolumn}
\usepackage{bm}
\usepackage{tikz}
\usetikzlibrary{math}
\usetikzlibrary{shapes.geometric, arrows}
\usetikzlibrary{positioning}
\usetikzlibrary{shapes,arrows}
\usetikzlibrary{intersections}
\usepackage{csvsimple}
\usepackage{changepage}
\usepackage[tbtags]{mathtools}
\usepackage{siunitx}
\usepackage{csquotes}

\usepackage{float}
\usepackage{subfigure}
\usepackage[
	nonumberlist, 				
	acronym,      				
	nomain,						
	nopostdot,					
]{glossaries}
\usepackage{glossary-superragged}
\usepackage[hidelinks]{hyperref}
\usepackage{color, colortbl}

\usepackage{wrapfig}

\usepackage{pgfplots}
\usepackage{tikz}
\usetikzlibrary{arrows}
\usepackage{amsmath}
\pgfplotsset{compat=newest}
\usepgfplotslibrary{fillbetween}
\usetikzlibrary{calc}
\def\centerarc[#1](#2)(#3:#4:#5)
{ \draw[#1] ($(#2)+({#5*cos(#3)},{#5*sin(#3)})$) arc (#3:#4:#5); }

\usepackage{amssymb}

\usepackage{nicefrac}

\usepackage{abstract}

\usepackage{multirow}
\usepackage{tabularx}
\usepackage{booktabs}
\usepackage{array}
\usepackage{longtable}
\newcolumntype{L}[1]{>{\raggedright\let\newline\\\arraybackslash\hspace{0pt}}m{#1}}
\newcolumntype{C}[1]{>{\centering\let\newline\\\arraybackslash\hspace{0pt}}m{#1}}
\newcolumntype{R}[1]{>{\raggedleft\let\newline\\\arraybackslash\hspace{0pt}}m{#1}}

\newacronym{3d}{3D}{three dimensional}
\newacronym{am}{AM}{additive manufacturing}
\newacronym{fdm}{FDM}{fused deposition modeling}
\newacronym{ism}{ISM}{in-space manufacturing}
\newacronym{iss}{ISS}{International Space Station}
\newacronym{fcb}{FCB}{Functional Cargo Block}
\newacronym{dem}{DEM}{discrete element method}
\newacronym{md}{MD}{molecular dynamics}
\newacronym{dc}{DC}{direct-current}
\newacronym[plural=PFCs,firstplural=parabolic flight campaigns (PFCs)]{pfc}{PFC}{Parabolic Flight Campaign}
\newacronym{fft}{FFT}{Fast Fourrier Transform}
\newacronym{cad}{CAD}{Computer Assisted Design}
\newacronym{ptfe}{PTFE}{polytetrafluoroethylene}
\newacronym{ps}{PS}{polystyrene}
\newacronym{nasa}{NASA}{National Aeronautics and Space Administration}
\newacronym{esamm}{ESAMM}{Extended Structure Additive Manufacturing Machine}
\newacronym{amf}{AMF}{Additive Manufacturing Facility}
\newacronym{us}{US}{United States}
\newacronym{usa}{USA}{United States of America}
\newacronym{bmgs}{BMGs}{Bulk Metallic Glasses}
\newacronym{esa}{ESA}{European Space Agency}
\newacronym{si}{SI}{International System of Units, abbreviated from French \textit{Syst\`{e}me International (d'unit\'{e}s)}}
\newacronym{dlr}{DLR}{German Aerospace Center}
\newacronym{liggghts}{LIGGGHTS}{\acrshort{lammps} Improved for General Granular and Granular Heat Transfer Simulations}
\newacronym{lammps}{LAMMPS}{Large-scale Atomic/Molecular Massively Parallel Simulator}
\newacronym{sjkr}{SJKR}{Simplified Johnson-Kendall-Roberts}
\newacronym{ded}{DED}{Directed Energy Deposition}
\newacronym{slm}{SLM}{Selective Laser Melting}
\newacronym{sls}{SLS}{Selective Laser Sintering}
\newacronym{eva}{EVA}{Extra-Vehicular Activity}
\newacronym{sem}{SEM}{Scanning Electron Microscopy}
\newacronym{RPM}{RPM}{Ramdom Positioning Machine}
\newacronym{rpm}{rpm}{revolutions per minute}
\newacronym{rise}{RISE}{Research Internships in Science and Engineering}
\newacronym{daad}{DAAD}{German Academic Exchange Service, abbreviated from German \textit{Deutscher Akademischer Austauschdienst}}
\newacronym{fsm}{FSM}{finite-state machine}
\newacronym{ir}{IR}{infrared}
\newacronym{pcbs}{PCBs}{Printed Circuit Boards}
\newacronym{pcb}{PCB}{Printed Circuit Board}
\newacronym{mcr}{MCR}{Modular Compact Rheometer}
\newacronym{sff}{SFF}{Solid Freeform Fabrication}
\newacronym{uv}{UV}{ultraviolet}
\newacronym{abs}{ABS}{acrylonitrile butadiene styrene}
\newacronym{hpde}{HPDE}{high density polyethylene}
\newacronym{pei}{PEI}{polyetherimide}
\newacronym{bff}{BFF}{BioFabrication Facility}
\newacronym{lens}{LENS}{Laser Engineered Net Shaping}
\newacronym{cnc}{CNC}{Computer Numerical Control}
\newacronym{ebf3}{EBF$^3$}{Electron Beam Free-Form Fabrication}
\newacronym{leo}{LEO}{Low Earth Orbit}
\newacronym{pc}{PC}{polycarbonate}
\newacronym{crissp}{CRISSP}{Customisable Recyclable International Space Station Packaging}
\newacronym{Athena}{Athena}{Advanced Telescope for High-ENergy Astrophysics}
\newacronym{lbm}{LBM}{Laser Beam Melting}
\newacronym{bam}{BAM}{Federal Institute for Materials Research and Testing, abbreviated from German \textit{Bundesanstalt f\"{u}r Materialforschung und-pr\"{u}fung}}
\newacronym{pbf}{PBF}{powder bed fusion}
\newacronym{eb}{EB}{Electron Beam}
\newacronym{2d}{2D}{two dimensional}
\newacronym{4d}{4D}{four dimensional}
\newacronym{ft4}{FT4}{Freeman Technology 4 Powder Rheometer}
\newacronym{dsc}{DSC}{Differential Scanning Calorimetry}
\newacronym{pmma}{PMMA}{polymethylmethacrylate}
\newacronym{1g}{$1g$}{gravity on-ground}
\newacronym{mug}{$\mu g$}{microgravity}
\newacronym{bcm}{BCM}{Box Counting Method}
\newacronym{mct}{MCT}{Mode Coupling Theory}
\newacronym{gmct}{gMCT}{granular Mode Coupling Theory}
\newacronym{itt}{ITT}{Integration Through Transients}
\newacronym{mfc}{MFC}{Mass Flow Controller}
\newacronym{ct}{CT}{computed tomography}
\newacronym{xct}{XCT}{X-ray computed tomography}
\newacronym{cv}{CV}{curriculum vitae}
\newacronym{pi}{PI}{principal investigator}
\newacronym{osp}{OSP}{orthogonal superimposed perturbation}
\newacronym{npi}{NPI}{Network Partnering Initiative}
\newacronym{ecsat}{ECSAT}{European Centre for Space Applications and Telecommunications}
\newacronym{eac}{EAC}{European Astronaut Centre}
\newacronym{estec}{ESTEC}{European Space Research and Technology Centre}
\newacronym{fps}{fps}{frames per second}
\newacronym{pdf}{pdf}{probability density function}
\newacronym{al}{Al}{aluminium}
\newacronym{ss}{\textit{SS}}{\textit{Smooth Surface}}
\newacronym{rs}{\textit{RS}}{\textit{Rough Surface}}
\newacronym{rcp}{rcp}{random close packing}
\newacronym{iop}{IoP UvA}{Institute of Physics of the University of Amsterdam}
\newacronym{mp}{MP}{Institute of Material Physics for Space}
\newacronym{elgra}{ELGRA}{European Low Gravity Research Association}
\newacronym{zarm}{ZARM}{Center of Applied Space Technology and Microgravity}
\newacronym{piv}{PIV}{particle image velocimetry}
\usepackage[framemethod=tikz]{mdframed}
\usepackage{lipsum}
\usepackage{enumitem}
\usepackage{mdframed}
\usepackage{physics}
\usepackage{setspace}

\usepackage[most]{tcolorbox}
\newtcolorbox{myBox}[3][]{
arc=2mm,
lower separated=true,
fonttitle=\bfseries,
colbacktitle=gray!10,
coltitle=black!50!black,
enhanced,
colframe=gray!10,
colback=gray!10,
title=#2,#1}

\usepackage{dirtytalk}
\usepackage{tcolorbox}
\usepackage[font=footnotesize]{caption}
\newtcolorbox{mybox}[1]{colback=green!6!white,colframe=black!75!black,fonttitle=\bfseries,title=#1}
\newtcolorbox{mybox2}{colback=red!5!white,colframe=red!75!black}

\usepackage{pifont}

\usepackage{soul,xcolor}
\setstcolor{red}

\usepackage{xcolor,hyperref}
\hypersetup{
   colorlinks,
   linkcolor={blue!50!black},
   citecolor={blue!50!black},
   urlcolor={blue!80!black}
} 

\definecolor{mycolor}{rgb}{0.122, 0.435, 0.698}

\usepackage{textgreek}

\newcommand{\Ubold}{\textbf{u}}
\newcommand{\del}[2]{\frac{\partial#1}{\partial#2}}

\title{{Elasto-viscoplastic Spreading: \\
\normalsize \large{\textit{from} Plastocapillarity \textit{to} Elastocapillarity}}}

\author[1,2]{Hugo L. Fran\c ca}
\author[2]{Maziyar Jalaal\footnote{m.jalaal@uva.nl, ORCID: 0000-0002-5654-8505}}
\author[3]{Cassio M. Oishi}

\affil[1] {Instituto de Ci\^encias Matem\'aticas e Computa\c{c}\~ao, Universidade de S\~ao Paulo, S\~ao Carlos, Brazil}
\affil[2]{
    Van der Waals-Zeeman Institute, Institute of Physics, University of Amsterdam, Amsterdam, The Netherlands}
\affil[3]{
	Departamento de Matem\'atica e Computa\c c\~ao, Faculdade de Ci\^encias e Tecnologia, Universidade Estadual Paulista ``J\'ulio de Mesquita Filho'', Presidente Prudente, Brazil}
\date{October 2021}

\begin{document}
\definecolor{brickred}{rgb}{0.8, 0.25, 0.33}
\definecolor{darkorange}{rgb}{1.0, 0.55, 0.0}
\definecolor{persiangreen}{rgb}{0.0, 0.65, 0.58}
\definecolor{persianindigo}{rgb}{0.2, 0.07, 0.48}
\definecolor{cadet}{rgb}{0.33, 0.41, 0.47}
\definecolor{turquoisegreen}{rgb}{0.63, 0.84, 0.71}
\definecolor{sandybrown}{rgb}{0.96, 0.64, 0.38}
\definecolor{blueblue}{rgb}{0.0, 0.2, 0.6}
\definecolor{ballblue}{rgb}{0.13, 0.67, 0.8}
\definecolor{greengreen}{rgb}{0.0, 0.5, 0.0}
\begingroup
\sffamily
\date{}
\maketitle
\endgroup

\begin{abstract}

We study the spreading of elastoviscoplastic (EVP) droplets under surface tension effects. The non-Newtonian material flows like a viscoelastic liquid above the yield stress and behaves like a viscoelastic solid below it. Hence, the droplet initially flows under surface tension forces but eventually reaches a final equilibrium shape when the stress everywhere inside the droplet falls below the resisting rheological stresses. We use numerical simulations and combine Volume-of-Fluid (VOF) method and an EVP constitutive model to systematically study the dynamics of spreading and the final shape of the droplets. 
The spreading process examined in this study finds applications in coating, droplet-based inkjet printing, and 3D printing, where complex fluids such as paints, thermoplastic filaments, or bio-inks are deposited onto surfaces. Additionally, the computational framework enables the study of a wide range of multiphase interfacial phenomena, from elastocapillarity to plastocapillarity. 


\textbf{keywords: Droplets $|$ Yield Stress Fluids $|$ Elasto-viscoplasticity $|$ Surface Tension $|$ Capillary Flows $|$ Plastocapillarity}

\end{abstract}

\section{Introduction}\label{sec:introduction}
The deposition and spreading of fluids over surfaces occur in a wide range of industrial applications, such as spray coating and printing~\cite{Barnes1998, Derby2010, Thompson2014, Mackay2018,lohse2022fundamental}. 
The fluids used in these applications often contain microscopic constituents like polymers or colloids, resulting in highly nonlinear macroscopic rheological features like elasticity, plasticity, and shear-dependent viscosity. 
Hence, understanding the rheological effects on the fluid mechanics of spreading is essential for further optimizing current systems or designing new materials for specific applications~\cite{ewoldt2021designing}. In the past few years, many studies have investigated the impact and spreading of droplets on surfaces for fluids with various rheological properties, including viscous fluids~\cite{tanner1979spreading,bonn2009wetting,Bergemann2018,Jalaal2019}, viscoelastic fluids~\cite{bergeron2000controlling,wang2015dynamic,izbassarov2016effects,wang2017impact,gorin2022universal}, and yield stress materials~\cite{Luu2009,saidi2010influence,saidi2011effects,blackwell2015sticking,Oishi2017,sen2020,Oishi2019,martouzet2021dynamic,d2022spreading,van2023viscoplastic}.
However, systematic experiments are generally complicated due to the high-dimensional interconnected parameter space. To this end, computer simulations can be employed to investigate the significance of rheological parameters independently from one another.

Our focus will be on a general class of nonlinear materials named elastoviscoplastic (EVP) fluids which can exhibit both elastic and plastic properties, as well as viscous behaviour. Many models of EVP fluids have been proposed previously. 
Saramito~\cite{Saramito2007, Saramito2009} presented a description of  EVP constitutive equations which combine classical viscoplastic models (Bingham or Herschel-Bulkley) with viscoelastic models (Oldroyd-B or PTT). Adopting a different perspective, de Souza Mendes \cite{Mendes2011} used a generalized viscoelastic model where material properties are functions of the strain rate in order to incorporate the yielding behaviour as well as the thixotropic effects. Dimitriou and McKinley~\cite{dimitriou2014comprehensive} included isotropic and kinematic hardening in their EVP formulation to capture various steady and unsteady flow responses. 
Recent reviews of the development and thorough comparison of different EVP models can be found in Fraggedakis et al.~\cite{Fraggedakis2016} and Saramito and Wachs~\cite{saramito2017progress}, where they comment both on the mathematical/physical properties of the models, as well as the numerical characteristics.
Moreover, many recent studies have incorporated the models above into computational frameworks to study various fluid dynamics problems such as flow around particles~\cite{Cheddadi2011,fraggedakis2016yielding,chaparian2020particle}, flow in channels~\cite{Nassar2011,izbassarov2021effect,kordalis2021investigation}, flow in porous media~\cite{de2018elastoviscoplastic,chaparian2020yield}, and flow around bubbles~\cite{de2019oscillations,moschopoulos2021concept}.
More related to the present investigations are the computational studies on droplets with EVP properties. Oishi \emph{et al.}~\cite{Oishi2017} studied the flow of materials on an inclined plane, considering elastic properties, where they captured the so-called avalanche effect where a decrease in viscosity (triggered by a stress field) induces a motion that successfully creates another decrease in viscosity. 
In follow-up studies,  Oishi \emph{et al.}~\cite{Oishi2019,oishi2019normal} have also included surface tension and thixotropic effects on the impact of EVP droplets on normal and inclined solid surfaces. 
Izbassarov and Tamissola \cite{Izbassarov2020} investigated how an EVP droplet deforms inside a Newtonian medium under simple shear and observe that the droplet deformation can present a non-monotonic behaviour as elasticity is changed. 


In the present work, our main goal is to study the spreading of EVP droplets on a surface. To this end, we aim to extend the previous numerical analysis~\cite{Jalaal2021} to consider not only viscoplastic rheology, but also the addition of elasticity. By using Saramito's EVP model~\cite{Saramito2007}, a parametric study will be carried over in order to understand how the addition of elasticity can affect the transient spreading dynamics and also the final shapes of elastoviscoplastic droplets. 

The paper is organized as follows. Section 2 presents a description of the problem, the governing equations used to model the flow and the rheology of the EVP material, and also an overview of the numerical method used in this work. In section 3, results are presented and various limits of the problem are visited. Section 4 concludes the results and presents future perspectives. Additional numerical details can be found in the appendices.

\section{Problem description and the Numerical framework}\label{sec:problem_description}

\subsection{Problem description: capillary spreading of a droplet}
We consider the spreading of an axisymmetric EVP droplet on a wetted surface. The choice of geometry has two main reasons. Firstly, in several applications, including 3D printing and spray coating, it is typical for droplets to spread and deposit on an existing layer of the same fluid~\cite{Bergemann2018, Jalaal2019}. Secondly, opting for a geometry that eliminates the presence of a triple contact line provides theoretical and computational advantages as it simplifies the associated complex physics related to boundary conditions and stress singularities.

\begin{figure}[!htb]
	\begin{center}
        \includegraphics[width=\linewidth]{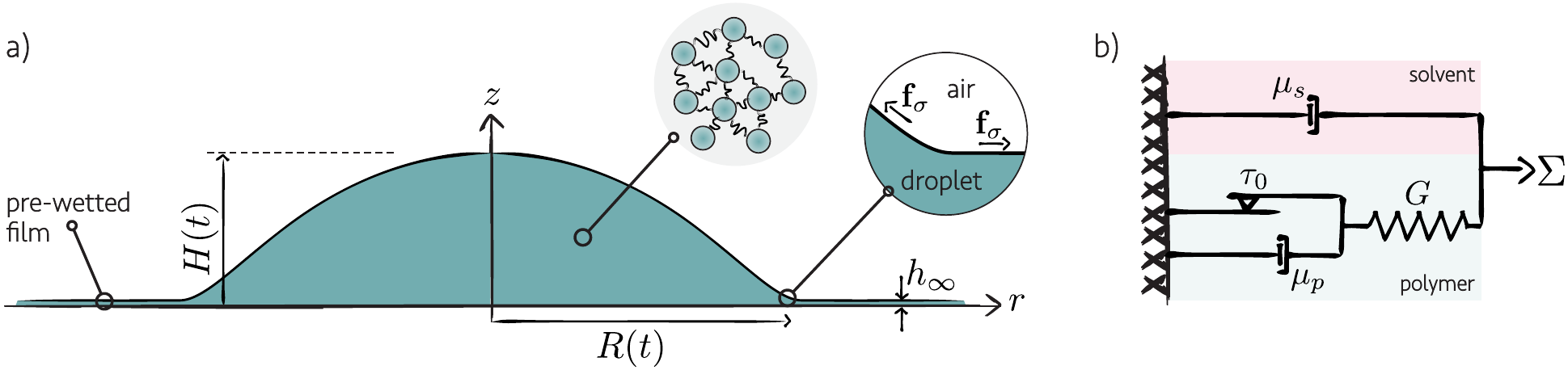}
		\caption{a) Sketch of the geometry of a spreading elastoviscoplastic droplet. We assume the droplet to be axisymmetric around the $z$-axis and spread on a thin pre-wetted film of the same material with thickness $h_{\infty}$. The radius and the height of the droplet are denoted respectively as $R(t)$ and $H(t)$.  Gravity is neglected, hence, the spreading is purely due to surface tension effects. The droplet includes microscopic constituents, resulting in macroscopic EVP models. b) Mechanical analog of Saramito's EVP model.}
		\label{fig:sketch_spreading}
	\end{center}
\end{figure}

%
As illustrated in Figure \ref{fig:sketch_spreading}a, a droplet is initially placed over a thin film of the same material and allowed to spread due to capillary
forces. Viscous forces oppose this spreading by slowing it down, and yield-stress is capable of stopping it completely as shown previously in \cite{Jalaal2021} for a purely viscoplastic scenario. The role of elasticity in this process, however, is less clear and a numerical framework will be implemented in order to understand such effects. To this end, we will model the rheology of the EVP material by the Saramito model~\cite{Saramito2007,Saramito2009}, which generalizes both the Bingham viscoplastic and the Oldroyd-B viscoelastic models. 
Figure \ref{fig:sketch_spreading}b shows a mechanical analog of this model. Below the yield stress the material behaves like a Kelvin-Voigt solid, while after yielding it flows as an Oldroyd-like fluid.

\subsection{Governing equations}
\label{sec:gov_equations}

The governing equations for the isothermal incompressible bi-phase flow, are the continuity and momentum conservation given by
\begin{align}
& \nabla \cdot \Ubold = 0, \label{navier1}\\
& \rho\left( \frac{\partial \Ubold}{\partial t} + \nabla \cdot ( \Ubold\Ubold ) \right)	= - \nabla p  + \nabla \cdot \bm{\tau} + \mathbf{f}_{g} + \mathbf{f}_{\sigma}, \label{navier0}
\end{align}
where $\Ubold$ and $p$ are the velocity and pressure fields. The gravitational force is defined as $\mathbf{f}_{g}=\rho \mathbf{g}$ where $\rho$ is the fluid mass density. In our numerical method, the surface tension force is also defined as a body force $\mathbf{f}_{\sigma} = \sigma \kappa \delta_s \textbf{n}$, where $\kappa$ is the curvature of the interface, $\sigma$ the constant surface tension coefficient, $\textbf{n}$ is the unit vector normal to the interface, and $\delta_s$ is the Dirac delta function centered on the interface~\cite{Tryggvason2011-book}.

The deviatoric stress tensor $\bm{\tau}$ is the sum of the solvent $\bm{\tau}^{s}$ and polymeric $\bm{\tau}^{p}$ contributions,
\begin{equation}
    \bm{\tau} = \bm{\tau}^{s} + \bm{\tau}^{p}.
    \label{decompExtraStress}
\end{equation}
	
The solvent stress contribution presents a Newtonian behaviour and the polymeric stress includes the memory effects, hence,
\begin{equation}
	\bm{\tau}^{s} = 2\mu_s\mathbf{D}, 
\label{tauS}
\end{equation}
where $\mu_s$ is the solvent viscosity and $\textbf{D} = \frac{1}{2}\left[ 
\nabla \textbf{u} + \left(\nabla \textbf{u}\right)^T \right]$ is the strain rate tensor. 
The polymer stress $\bm{\tau}^{p}$ includes the elasto-viscoplastic contribution, emerging from the microstructures. For viscoelastic fluids, it is common to model the contribution of the polymeric stress as $\bm{\tau}^{p} = \frac{\mu_p}{\lambda} f(\mathbf{A})$, where $\lambda$ is the relaxation time, and $f(\mathbf{A})$ is a strain function of the conformation tensor $\mathbf{A}$~\cite{oldroyd1950formulation,stone2023note,Snoeijer2020}.
We assume $f(\mathbf{A}) = (\mathbf{A} - \mathbf{I})$ and $\mathbf{A}$ follows a liner relaxation law $\stackrel{\raisebox{.3ex}{ \hskip -.1cm$\scriptscriptstyle{\bigtriangledown}$}}{\mathbf{A}} = - \frac{1}{\lambda} (\mathbf{A} -\mathbf{I})$, where $\stackrel{\raisebox{.3ex}{ \hskip -.1cm$\scriptscriptstyle{\bigtriangledown}$}}{(\cdot)}$ is the upper-convected time derivative. Adding a critical yield stress at which the material switches between liquid and solid phase, we arrive at a constitute law, 
first formulated by Saramito~\cite{Saramito2007,Saramito2009}. The model combines the Bingham (viscoplastic)~\cite{bingham1917investigation,frigaard2019simple} and the upper convected  Maxwell (viscoelastic) models~\cite{oldroyd1950formulation, morozov2015introduction,Snoeijer2020, hinch2021oldroyd, stone2023note}. The polymeric stress then is governed by,
\begin{equation}
	 \lambda \stackrel{\raisebox{.3ex}{ \hskip -.3cm$\scriptscriptstyle{\bigtriangledown}$}}{\bm{\tau}^p} + \max{\left(0, 1 - \frac{\tau_0}{\norm{\bm{\tau}^p}}\right)}\bm{\tau}^p = 2\mu_p\mathbf{D},
\label{tauP}
\end{equation}
 where, $\tau_0$ is the yield stress, $\norm{\bm{\tau^p}} = \sqrt{\text{tr}({\bm{\tau}^p}^2)}$ 
, and $\stackrel{\raisebox{.3ex}{ \hskip -.3cm$\scriptscriptstyle{\bigtriangledown}$}}{\bm{\tau}^p}$ is given by
\begin{equation}
	\stackrel{\raisebox{.3ex}{ \hskip -.3cm$\scriptscriptstyle{\bigtriangledown}$}}{\bm{\tau}^p}
	=
	\del{\bm{\tau}^p}{t} + \left( \Ubold\cdot\nabla \right)\bm{\tau}^p - (\nabla\Ubold)\bm{\tau}^p - \bm{\tau}^p(\nabla\Ubold)^T.
\label{upperConvected}
\end{equation}

A dimensionless version of these equations can be obtained by scaling the variables as follows
\begin{equation}
	{ \textbf{x} = \mathcal{L}\bar{\textbf{x}}, \hspace{10pt} t = \frac{\mathcal{L}}{U}\bar{t}, \hspace{10pt} \Ubold = U\bar{\Ubold}, \hspace{10pt} p =  \rho_d\,U^2\bar{p}, \hspace{10pt} \bm{\tau}^p = \rho_d\,U^2\bar{\bm{\tau}}^p },
\label{adimensionalizacao}
\end{equation}
 where  $\bm{x}$ is the position vector, $t$ is time, $U = \sqrt{\frac{\sigma}{\rho_d \mathcal{L}}}$ is the characteristic velocity, $\rho_d$ is the droplet density, 
and the length scale is $\mathcal{L} = \left[3\mathcal{V}/(4\pi)\right]^{1/3}$ with $\mathcal{V}$ being the volume of the droplet ($\mathcal{L}$ can be seen as the radius of a corresponding spherical droplet with same volume $\mathcal{V}$).

Removing, for convenience, the bars in \eqref{adimensionalizacao}, the dimensionless governing equations for the droplet phase are given by
\begin{align}
& \nabla \cdot \Ubold = 0, \label{eqAdim1}\\
& { \frac{\partial \Ubold}{\partial t} + \nabla \cdot ( \Ubold\Ubold ) = - \nabla p  + \nabla \cdot \left( 2\ Oh_s  \ \textbf{D} \right) + \nabla\cdot \bm{\tau}^p + Bo\cdot \textbf{g} + \kappa \delta_s\textbf{n}, } \label{eqAdim2}\\
& De \stackrel{\raisebox{.3ex}{ \hskip -.3cm$\scriptscriptstyle{\bigtriangledown}$}}{\bm{\tau}^p} + \max{\left(0, 1 - \frac{\mathcal{J}}{\norm{\bm{\tau}^p}} \right)}\bm{\tau}^p = 2\ Oh_p\ \textbf{D}
		\label{eqAdim3}
\end{align}

with the dimensionless groups: Ohnesorge numbers ($Oh_s$ and $Oh_p$), Bond number ($Bo$), plastocapillary number ($\mathcal{J}$), and Deborah number ($De$) 
defined respectively as
\begin{equation}
	Oh_s = \frac{\mu_s}{\sqrt{\rho_d \sigma \mathcal{L}}}, \hspace{15pt} Oh_p = \frac{\mu_p}{\sqrt{\rho_d \sigma \mathcal{L}}}, \hspace{15pt} Bo = \frac{\rho_d g \mathcal{L}^2}{\sigma}, \hspace{15pt} \mathcal{J} = \frac{\tau_0 \mathcal{L}}{\sigma}, \hspace{15pt}De = \lambda\sqrt{\frac{\sigma}{\rho_d \mathcal{L}^3}}, \hspace{15pt}
\label{paramAdimensionais}
\end{equation}
The Ohnesorge numbers compare the characteristic Rayleigh timescale and the visco-capillary time scale and, for this problem, could be seen as normalized solvent or polymeric viscosities. The Bond number compares the gravitational stresses and capillary pressure. In the present study, we focus on pure capillary spreading, where gravity is negligible, hence $Bo=0$ in all simulations. Although we admit that the limit of large $Bo$ and negligible surface tension is interesting and important for geophysical problems like lava flows, mudslides, and snow avalanches where large-scale EVP fluids flow under gravity~\cite{balmforth2001geophysical,ancey2007plasticity}. The plastocapillary number compares the yield stress and the capillary stresses, and the Deborah number is the ratio of the polymeric relaxation time to the characteristic time-scale of the problem. Note that many authors choose to replace parameters $Oh_s$ and $Oh_p$ by $Oh = Oh_s + Oh_p$ and $\beta = Oh_s / (Oh_s+Oh_p)$ ~\cite{Viezel2020}, where $\beta$ can be seen as the normalized ratio of solvent to apparent viscosity. We also note that, using the present multiphase method, we also have fluid motion in the air phase. Therefore, a set of equations similar to equations \eqref{eqAdim1}-\eqref{eqAdim3}, but with Newtonian rheology, is also solved for the air flow. Consequently, two other nondimensional parameters are also relevant: the density ratio between droplet and air $(\rho_d/\rho_a)$ and viscosity ratio ($\mu_s/\mu_a$), where subscript $a$ denotes air properties. Both of these ratios are kept constant with a value of 100. 

%

\noindent Limits of equations \eqref{eqAdim1}-\eqref{eqAdim3} worth mentioning:

\begin{myBox}[breakable, enhanced]{}


\textbf{[Case 1:} $\mathcal{J} = 0$ \textbf{--- Visco-elastocapillarity Regime]} \\
Omitting the yield stress leads to a purely viscoelastic (Oldroyd-B) fluid under capillary forces~\cite{mckinley2005visco,clasen2006beads,bhat2010formation,ardekani2010dynamics,deblais2018pearling}. 

\vspace{7pt}
\textbf{[Case 2:} $De = 0$ \textbf{--- Plastocapillarity Regime]}\\
Excluding viscoelastic (memory) effects results in a purely viscoplastic (Bingham) free surface flow under capillary stresses~\cite{jalaal2023plastocapillarity}. Note that, substituting $De = 0$ in equation \eqref{eqAdim3}, from equation \eqref{eqAdim2}, we arrive at the deviatoric stress of
    \begin{equation}
        \bm{\tau} = 2\left[ Oh_s + Oh_p + \frac{\mathcal{J}}{2\norm{\textbf{D}}} \right]\textbf{D}
    \label{eq:vp_constitutive1}
    \end{equation}

for yielded regions. For a Bingham fluid, $\textbf{D} = 0$ when the material is unyielded. Hence, numerical regularization is required. In practice, for this limit, instead of the EVP constitutive model, we will solve a regularized version of the generalized Newtonian model above~\cite{Jalaal2021, sanjay2021bursting}.

\vspace{7pt}
\textbf{[Case 3:} $\mathcal{J} = 0$ and $De = 0$  \textbf{--- Newtonian Regime I]}\\
Without elasto-plastic rheology, we simply have a Newtonian fluid spreading due to surface tension, and the deviatoric stress tensor is
     \begin{equation}
        \bm{\tau} = 2 \ (Oh_s + Oh_p) \ \textbf{D}.
        \label{eq:newt_constitutive}
    \end{equation}

\vspace{7pt}
\textbf{[Case 4:} $De \rightarrow \infty$  \textbf{--- Newtonian Regime II]}\\
For finite $Oh_p$ and $\mathcal{J}$, if $De \rightarrow \infty$, we arrive at $\bm{\tau}^p \rightarrow 0$ and $\stackrel{\raisebox{.3ex}{ \hskip -.3cm$\scriptscriptstyle{\bigtriangledown}$}}{\bm{\tau}^p} \rightarrow 0$. 
Hence, we will have another Newtonian spreading with
     \begin{equation}
        \bm{\tau} \approx 2 \ Oh_s \ \textbf{D}.
        \label{eq:newt_constitutive}
    \end{equation}
In other words, the polymers behave like passive scalars and do not contribute to the dynamics of the problem. This regime might not be physical because, in reality, the viscous and plastic dissipations might also depend on the relaxation time. Note, if $Oh_p/De$ ($\sim$ elastic module) remained finite, then equation \eqref{eqAdim3} converged to the constitutive model of an elastic solid~\cite{Snoeijer2020}.

\vspace{7pt}
\textbf{[Case 5:} $\mathcal{J} \rightarrow \infty$  \textbf{--- Elastocapillarity Regime]}\\
For a finite $De$, when the plastocapillary number is large, equation \eqref{eqAdim3} reduces to 
     \begin{equation}
        \stackrel{\raisebox{.3ex}{ \hskip -.3cm$\scriptscriptstyle{\bigtriangledown}$}}{\bm{\tau}^p} = 2\frac{Oh_p}{De} \mathbf{D},
        \label{eq:KV}
    \end{equation}

\emph{i.e.,} the stress inside the material remains below the yield stress and the polymeric response is elastic. In fact, $Oh_p/De  = \mu \mathcal{L} / \sigma \lambda$ is the elastocapillary number (or the inverse of it~\cite{mckinley2005visco}). In this limit, the droplet behaves like a Kelvin-Voigt solid that spreads under capillary stress~\cite{kelvin1865elasticity, voigt1892ueber,lakes1998viscoelastic}. This regime is highly related to other soft wetting phenomena, where capillarity forces deform soft solids~\cite{style2017elastocapillarity,bico2018elastocapillarity,andreotti2020statics}.
\end{myBox}

In all of the cases above, the system of equations is closed with appropriate boundary conditions. We apply a no-slip condition on the solid wall. Rotational symmetry is applied at the center of the droplet and outflow boundary conditions are used at the two boundaries distant from the droplet.

\subsection{Initial condition \& numerical method}
\label{sec:general_numerical}

We used the open source code Basilisk C to solve the equations described in the previous section.
An overview of the numerical procedure will be given in this section, but detailed descriptions of Basilisk can be viewed in \cite{Popinet2013-Basilisk} (also see appendices \ref{appA} and \ref{appB}).


The simulation is setup by initializing a droplet according to the following shape
\begin{equation}
    h(r, 0) = h_{\infty} + R_0\max{\left(0, 1 - (r/R_0)^2\right)}, \hspace{20pt} \text{ with }\hspace{20pt}  R_0 = 1,
\label{eq:initial_shape}
\end{equation}
which represents a half-parabola in an axisymmetric coordinate system. The droplet is placed in a squared domain with dimensions $[0, 5R_0] \times [0, 5R_0]$ which is fully discretized with a non-uniform quadtree grid~\cite{popinet2003gerris,popinet2009accurate}. To accurately resolve the flow structure inside the droplet and its shape, we apply increased refinement levels for the liquid phase and also at the interface (see appendix \ref{appA}). As the droplet spreads over time, the mesh is also adapted so that the refined region follows the interface.

The interface is represented implicitly by the Volume of Fluid (VOF) scheme \cite{Hirt1981}, in which each mesh cell stores a value representing the fraction of droplet fluid. Density and viscosity are then locally determined based on the volume fraction $c(\textbf{x}, t)$ according to
\begin{align}
    \rho(c) = & \  c \ \rho_d + (1 - c)\rho_a, \\
    \mu(c) = & \ c \  \mu_d + (1 - c)\mu_a.
\label{eq:density_viscosity}
\end{align}
where $\rho$ and $\mu$ represent the density and dynamic viscosity of a fluid, respectively. The subscripts $d$ and $a$ represent, the droplet and the air, respectively.

This volume fraction field $c$ is then advected over time by solving the equation
\begin{equation}
    \del{c}{t} + \nabla \cdot (c \textbf{u}) = 0.
\label{eq:vof_advection}
\end{equation}

The numerical code then solves the governing equations using a projection method and a multilevel Poisson solver (see \cite{popinet2009accurate,Popinet2015} for more details of the VOF implementation).

\section{Results and discussion}
We construct the discussion on EVP spreading by first analyzing the results of pure viscoplastic and viscoelastic droplets. In all simulations in this section, the following parameters are fixed: $Oh_s = 1/90$, $Oh_p = 8/90$, $Bo = 0$.
\subsection{Plastocapillarity}
\label{sec:vpscaling}
Theoretical, computational, and experimental results for viscoplastic droplet spreading ($De = 0$) have been previously presented~\cite{Jalaal2021, van2023viscoplastic}. We will revisit this limit first to validate the simulation and also to extend the available results for higher plastocapillary numbers. To this end, we will focus on the properties of the final shape of the droplet when $t\rightarrow \infty$. Unlike Newtonian droplets, droplets of yield-stress fluids will be arrested at this limit, resulting in a finite final radius $R_f$ and height $H_f$. In a pure viscoplastic case, these features can be explained by balancing the capillary stress and yield stress~\cite{Jalaal2021}, resulting in theoretical scaling laws: $R_f/\mathcal{L} = 1.74\,\mathcal{J}^{-1/7}$, $H_f/\mathcal{L} = 1.09\,\mathcal{J}^{2/7}$. Note, the pre-factors are obtained from asymptotic analysis~\cite{Jalaal2021}. Also note, to test our numerical results against these laws, we must choose a stoppage criteria in our simulations since we use a regularized model in the viscoplastic limit. We do this based on the nondimensional kinetic energy ($E_k= (1/\sigma \mathcal{L}^2)\int_V{\frac{1}{2} \ \rho_d \ \norm{\textbf{u}}^2 \ dV}$) of the droplet, such that the simulation is stopped when $E_k <10^{-6}$.

\begin{figure}[h!]
    \centering
    \includegraphics[width=0.5\textwidth]{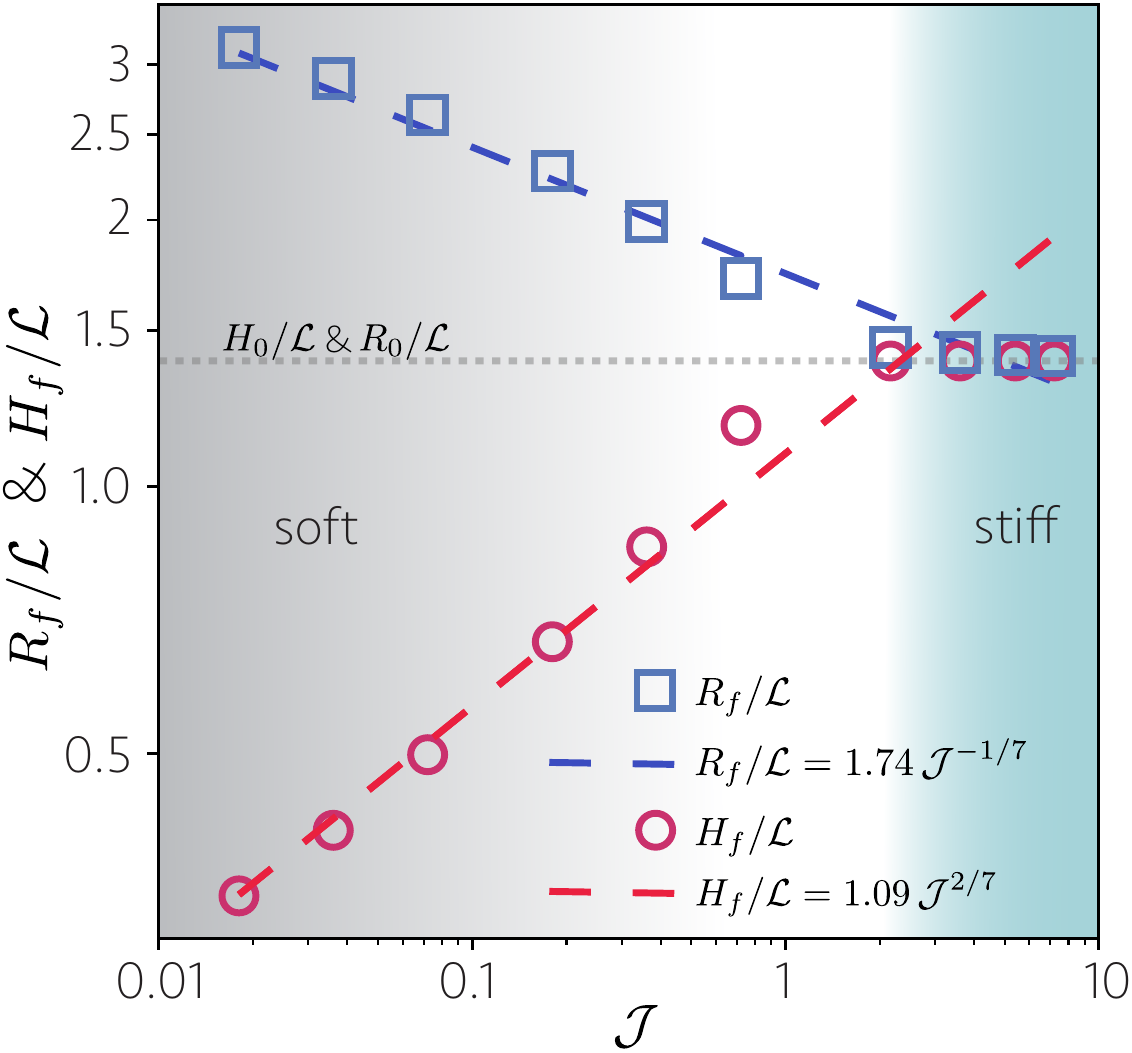}
    \caption{Droplet final radius and height as a function of the plastocapillary number. Symbols are the present numerical simulations. The thick dashed lines show the theoretical predictions from \cite{Jalaal2021}. The gray horizontal line indicates the initial radius and height of the droplets.}
    \label{fig:vp_valid_mazi_theory}
\end{figure}

 Figure \ref{fig:vp_valid_mazi_theory} shows the final radius and height of our simulated droplets versus the plastocapillary number $\mathcal{J}$. As expected, due to the higher influence of yield stress on capillary-driven spreading, the final radius decreases with $\mathcal{J}$, while the final height increases. Both final radius and height reach a plateau after a certain value of plastocapillary number, $\mathcal{J} \approx 1$ (see Video I). At this limit, the droplets practically get stuck at their initial shape, since the capillary stress is not strong enough to overcome the yield-stress at all.
 There exists a transition regime between the low yield-stress scaling laws (soft) and the high yield-stress plateau (stiff) regimes, where the numerical results smoothly vary between the two. There is a good agreement between the theory and numerical results when the droplets are soft enough to yield. As plastocapillary number increases, the droplet does not entirely yield and hence break the assumption in the theoretical predictions. Note that the exact point of transition from soft to stiff limit depends on the initial conditions. This could have important implications for some applications such as 3D printing~\cite{rauzan2018particle,friedrich2020corner,milazzo20233d,van2023viscoplastic,garcia2023fourier,saadi2022direct}, when depending on the values of $\mathcal{J}$, the history (shape) of droplet/filament at the time of deposition can influence the final geometry of the print. 

\subsection{Visco-elastocapillarity}
\label{section:ve_case}
\begin{figure}[h]
    \centering
    \includegraphics[width=0.8\textwidth]{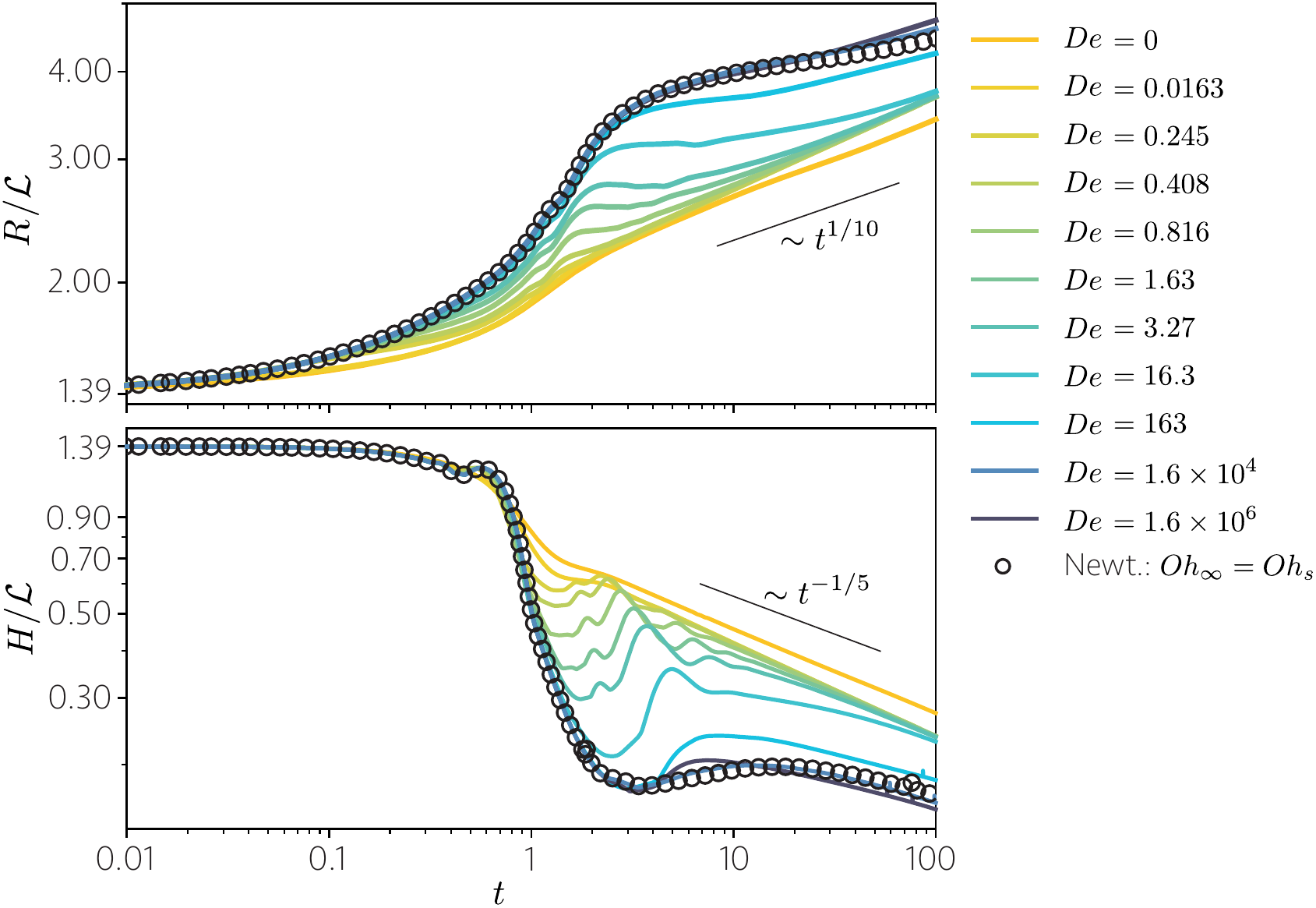}
    \caption{Spreading radius (top) and height (bottom) over time for different values of the Deborah number.}
    \label{fig:ve_radiustime}
\end{figure}

We continue by considering the case of viscoelastic materials without plasticity, that is, $\mathcal{J} = 0$. In this situation, the constitutive model reduces to an Oldroyd-B fluid. Figure \ref{fig:ve_radiustime} shows the droplet radius (top) and height (bottom) over time for different values of $De$ (see Video II).
In the Newtonian limit ($De = 0$), the spreading eventually follows the rate predicted by Tanner's law \cite{tanner1979spreading}, \emph{i.e.}, $R \propto t^{1/10}$ and $H \propto t^{-1/5}$.
For the viscoelastic case, we note that, in the first moments, the droplet spreads considerably more as we increase $De$. We anticipate this is due to the increased relaxation time of the fluid.
As we increase the relaxation time (or $De$), the stresses take a longer time to develop as the flow field develops inside the droplet, consequently, the droplet spreads more since the internal stress is smaller during this transient period. As a consequence, interestingly, the spreading curves converge to an apparent Newtonian limit when $De \rightarrow \infty$. In this limit, the polymeric stress does not have enough time to develop at all within the timescale of the simulation, and we actually recover a Newtonian droplet that only exhibits the solvent stress, \emph{i.e.}, we have a Newtonian fluid with Ohnesorge number $Oh_{\infty} = Oh_s$ (circles in figure~\ref{fig:ve_radiustime}). For the intermediate values of $De$, the interface experiences an oscillatory behaviour (see figure \ref{fig:ve_radiustime}), where the droplet height first reaches a local minimum and then increases again as elastic stresses build up, eventually reaching a decaying regime. 

\begin{figure}[htb!]
    \centering
    \includegraphics[width=0.8\textwidth]{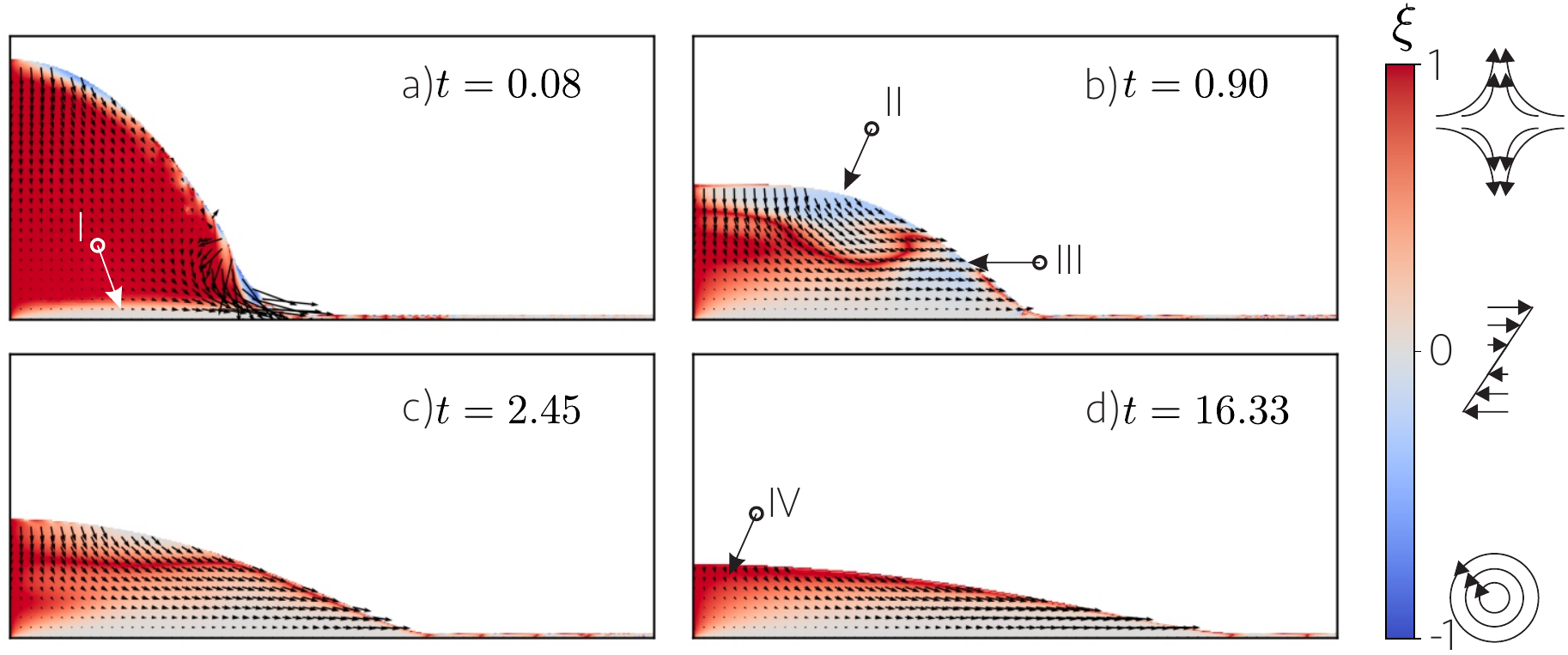}
    \caption{Flow structure inside a viscoelastic droplet with $De = 0.245$ at different time stamps. Regions of $\xi = 0$ indicate purely shear flow, $\xi=1$ extensional flow, and $\xi = -1$ rotation.}
    \label{fig:ve_flowtype}
\end{figure}

To further analyse the anatomy of viscoelastic spreading, we inspect the flow field inside the droplet using a flow parameter \cite{Afonso2011}:
\begin{equation}
    \xi = \frac{|\textbf{D}| - |\mathbf{\Omega}|}{|\textbf{D}| + |\mathbf{\Omega}|},
\label{eq:flow_type}
\end{equation}
where $\textbf{D}$ and $\mathbf{\Omega}$ are the deformation rate and vorticity tensors (the symmetric and antisymmetric components of $\nabla \mathbf{u}$), respectively: $ \textbf{D} = 1/2\ \left[ \nabla \textbf{u} + 
    (\nabla\textbf{u})^T \right]$ and $\mathbf{\Omega} = 1/2\left[ \nabla \textbf{u} - 
    (\nabla\textbf{u})^T \right]$.
%
The flow parameter can vary in the range $\xi \in [-1, 1]$, where $\xi = 1$ represents a purely extensional flow, $\xi = 0$ shear flow and $\xi = -1$ indicates solid-like rotation. Figure \ref{fig:ve_flowtype} shows the value of $\xi$ inside a viscoelastic droplet with $De = 0.245$ for different time frames. At the beginning of spreading, the flow close to the axi-symmetry axis and in the bulk of the droplet is predominantly extensional (red color).
Meanwhile, a shear-dominant boundary layer forms from the contact line and across the substrate (zone I figure~\ref{fig:ve_flowtype}a). Two rotating regions also develop close to the interface and grow over time (zones II and III in figure~\ref{fig:ve_flowtype}b) but decay as the droplet further spreads. Eventually, the flow inside the droplet is mainly extensional in the centre (zone IV in figure~\ref{fig:ve_flowtype}d) and shear-dominant close to the substrate.



In the next step, we will study the additional effects of plasticity on the complex flow structures shown above in an EVP spreading.


\subsection{Elastoviscoplastic spreading}
\begin{figure}[h]
    \centering
    \includegraphics[width=0.8\textwidth]{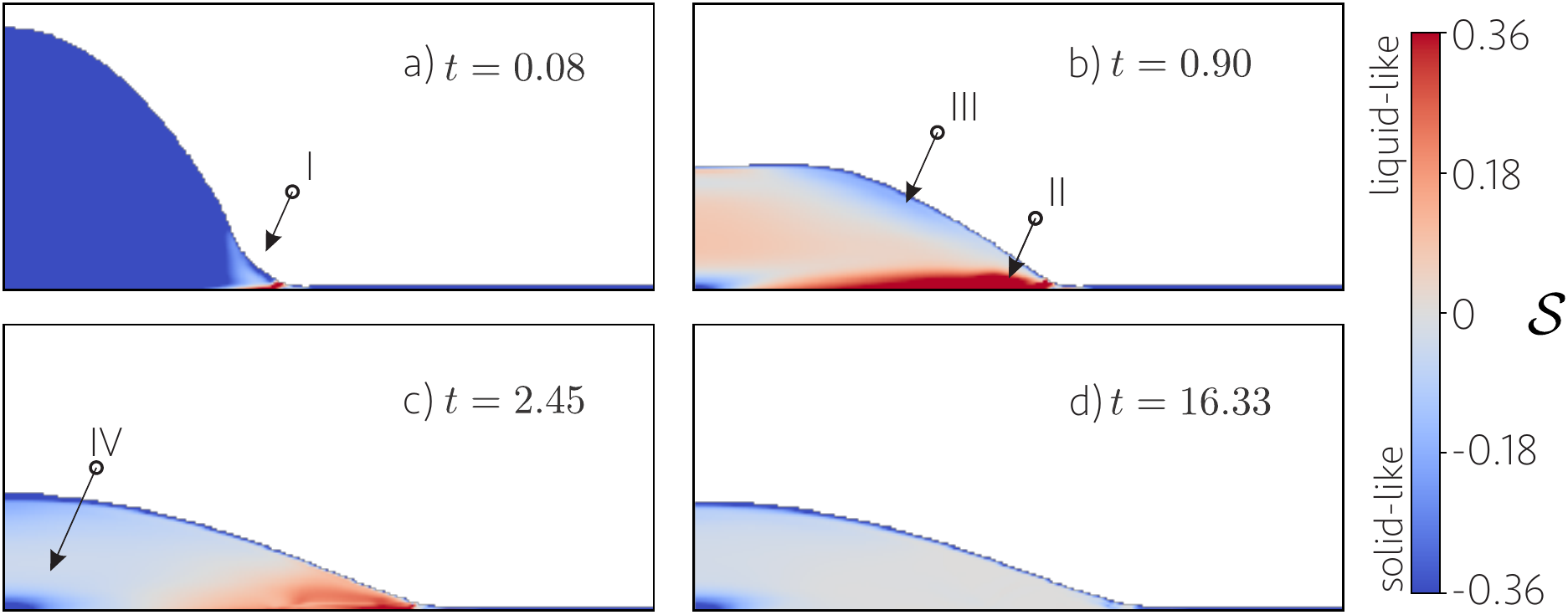}
    \caption{Distribution of $\mathcal{S}$ at different timestamps inside a elastoviscoplastic droplet with $De = 0.816$ and $\mathcal{J} = 0.18$. Blue regions indicate stress below the plastocapillary number $\mathcal{J}$ (unyielded) and red regions are above $\mathcal{J}$ (yielded). We note that the value of $\mathcal{J}$ used to calculate $\mathcal{S}$ is different in each row of this figure.} 
    \label{fig:evp_tau_time}
\end{figure}

We now consider the general scenario of elastoviscoplastic spreading, \emph{i.e}, $De$ and $\mathcal{J}$ are both finite. In this case, all the nonlinear properties mentioned earlier (including the coexistence of solid and liquid states in plastic fluids and the time-dependent characteristics of viscoelastic fluids) are simultaneously observed.
We inspect the nondimensional polymeric stress $\bm{\tau}^{p}$ and observe in which regions this stress is above (yielded) or below (unyielded) the value of $\mathcal{J}$ (see \cite{Izbassarov2020}). 
%
Figure \ref{fig:evp_tau_time} shows the value of scalar $\mathcal{S} = \log(\norm{\bm{\tau}^{p}}) - \log{(\mathcal{J})}$ at different times inside a droplet with $\mathcal{J} = 0.18$ and $De = 0.816$. $\mathcal{S}>0$ means the material is fluidized and flows like a viscoelastic fluid, and $\mathcal{S}<0$ means the material is not yielded and behaves like a viscoelastic solid. As expected, the droplet is initially mostly unyielded (large blue region) since we assume there is no internal stress as our initial condition. The stress begins to increase from the contact line, which is the location of the highest curvature (zone I in figure~\ref{fig:evp_tau_time}a). After some time, most of the droplet is yielded (red colors) with a particularly higher stress region near the wall and droplet edge (zone II in figure~\ref{fig:evp_tau_time}b). Meanwhile, a moving plug forms near the interface of the droplet (zone III in figure~\ref{fig:evp_tau_time}b), correlated with the rotating regions shown in figure~\ref{fig:ve_flowtype}. As time continues to increase, the droplet solidifies once again with a static plug in the centre (zone IV in figure~\ref{fig:evp_tau_time}c), leading to a full stoppage. Note that, unlike the pure viscoplastic case, no regularization is needed in the case of EVP materials; the viscoelastic solid will reach an equilibrium state as long as stress everywhere inside the droplet is below the yield stress.

\begin{figure}[hbt!]
    \centering
\includegraphics[width=0.85\textwidth]{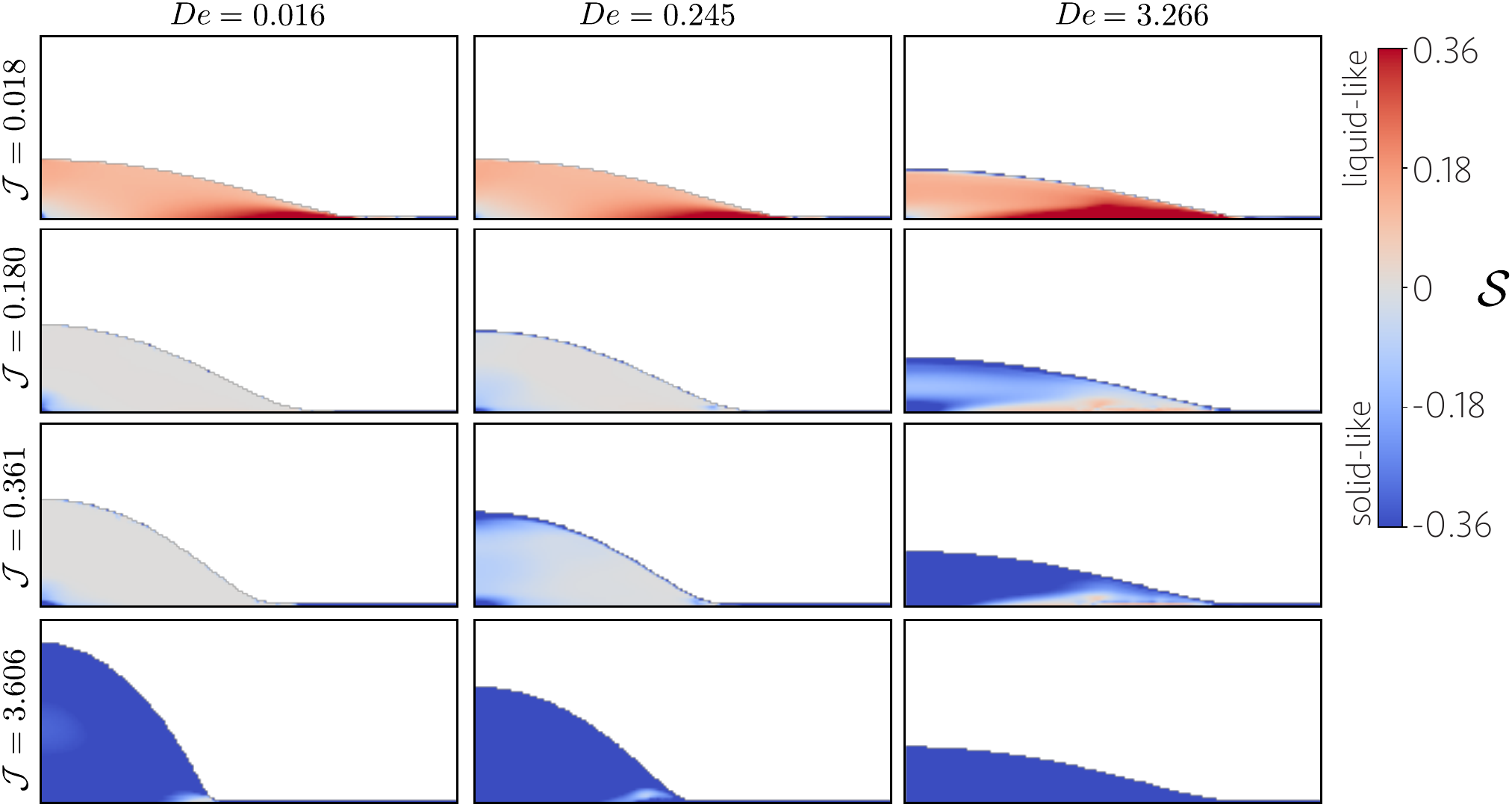}
    \caption{Distribution of $\mathcal{S}$ inside the elastoviscoplastic droplet for different values of $De$ and $\mathcal{J}$. All snapshots are taken at time $t = 8.16$ (see Video III). Blue regions indicate stress below the plastocapillary number $\mathcal{J}$ (unyielded) and red regions are above $\mathcal{J}$ (yielded). We note that the colorbar limits do not include the total range of values present in the data, since we are only interested in visualizing regions that are above or below zero .}
    \label{fig:evp_compareTau_Wi_J}
\end{figure}

We systematically extended the analysis above by changing the control non-dimensional parameters. 
Figure \ref{fig:evp_compareTau_Wi_J} demonstrates the value of $\mathcal{S}$
for different combinations of $\mathcal{J}$ and $De$ (see Video III).
For a given $De$, increasing the value of $\mathcal{J}$ leads to larger unyielded regions inside the droplet, which is expected as we increase the material yield-stress. Hence, for a given $De$, and at a given time, the droplet spreads less as the values of $\mathcal{J}$ increase. 
For a given $\mathcal{J}$, increasing the Deborah number also generates larger unyielded regions (in blue). Similarly to the explanation in section \ref{section:ve_case}, increasing $De$
results in smaller values of  $\bm{\tau}^{p}$ and more regions below the yield-stress. 
Intriguingly, at the high $\mathcal{J}$ cases (last row in figure \ref{fig:evp_compareTau_Wi_J}), it is notable to observe that by increasing $De$, the droplet spreads significantly even though it almost entirely behaves like a solid. This is because the current EVP formulation models the unyielded case as a viscolastic solid, allowing for a finite deformation rate (as opposed to a Bingham rigid solid where the deformation rate is zero for the solid state). This means that the material can experience deformation even below the yield stress, as seen clearly in the examples when both $\mathcal{J}$ and $De$ are high (elastocapillarity regime).

\begin{figure}[hbt!]
    \centering
\includegraphics[width=0.8\textwidth]{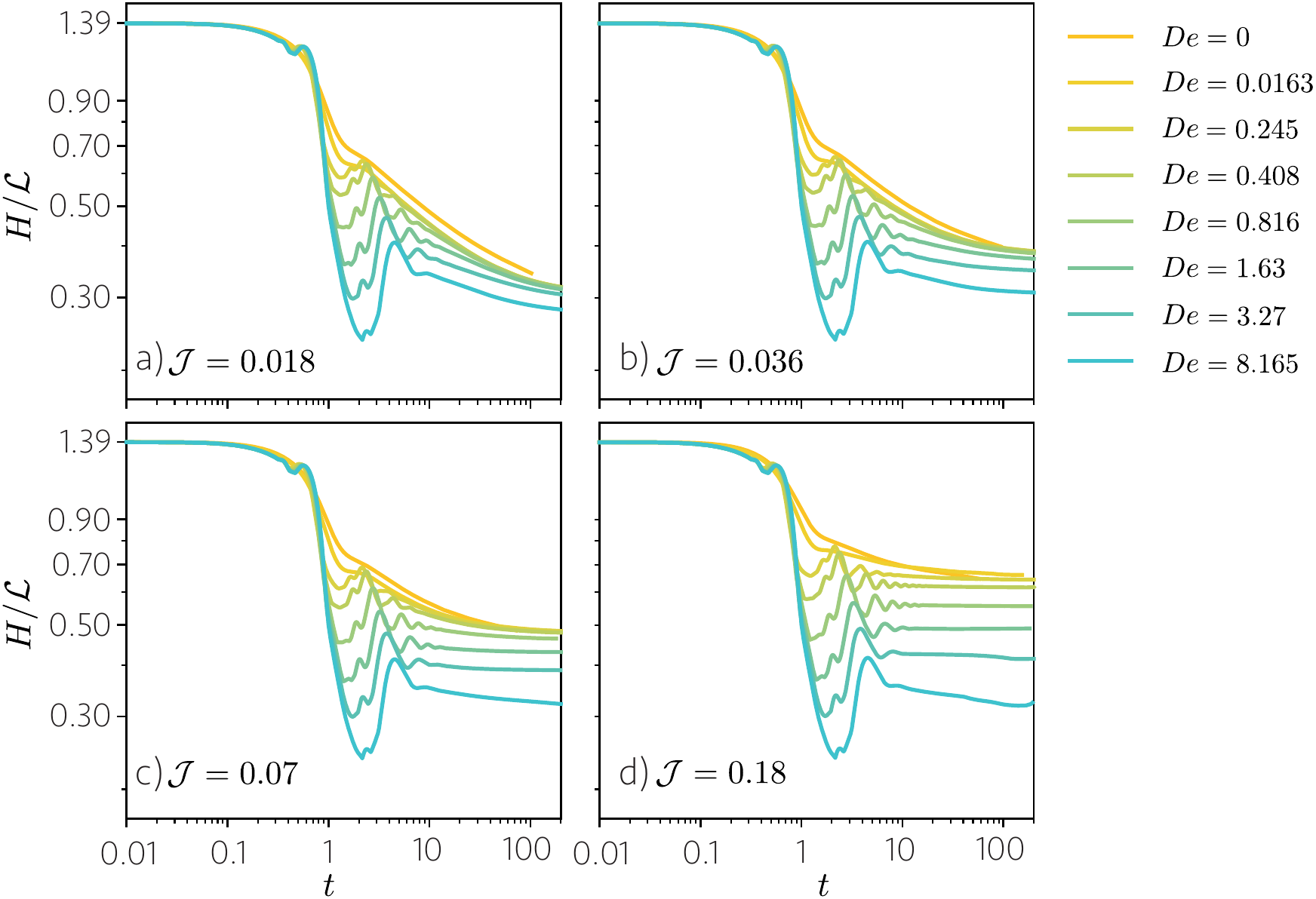}
    \caption{The evolution of droplet height as a function of the plastocapillary and Deborah numbers.}
    \label{fig:evp_heighttime}
\end{figure}


For a better quantitative analysis of the dynamics shown above, we look at the variation of droplet height over time for different values of $\mathcal{J}$ and $De$, as shown in  Figure \ref{fig:evp_heighttime}. For EVP droplets
the spreading stops and a final shape is reached. This is similar to the plastocapillarity regime, however, the elasticity clearly influences the final shape. Similarities to the Visco-elastocapillarity regime can also be seen, particularly in the local overshoot and the oscillation of the height that is observed for most cases and amplifies with $De$. At the same time, the impact of $De$ on the dynamics is more pronounced for higher values of $\mathcal{J}$, resulting in a larger difference in the final shape (height and radius). 

\begin{figure}[hbt!]
    \centering
\includegraphics[width=0.6\textwidth]{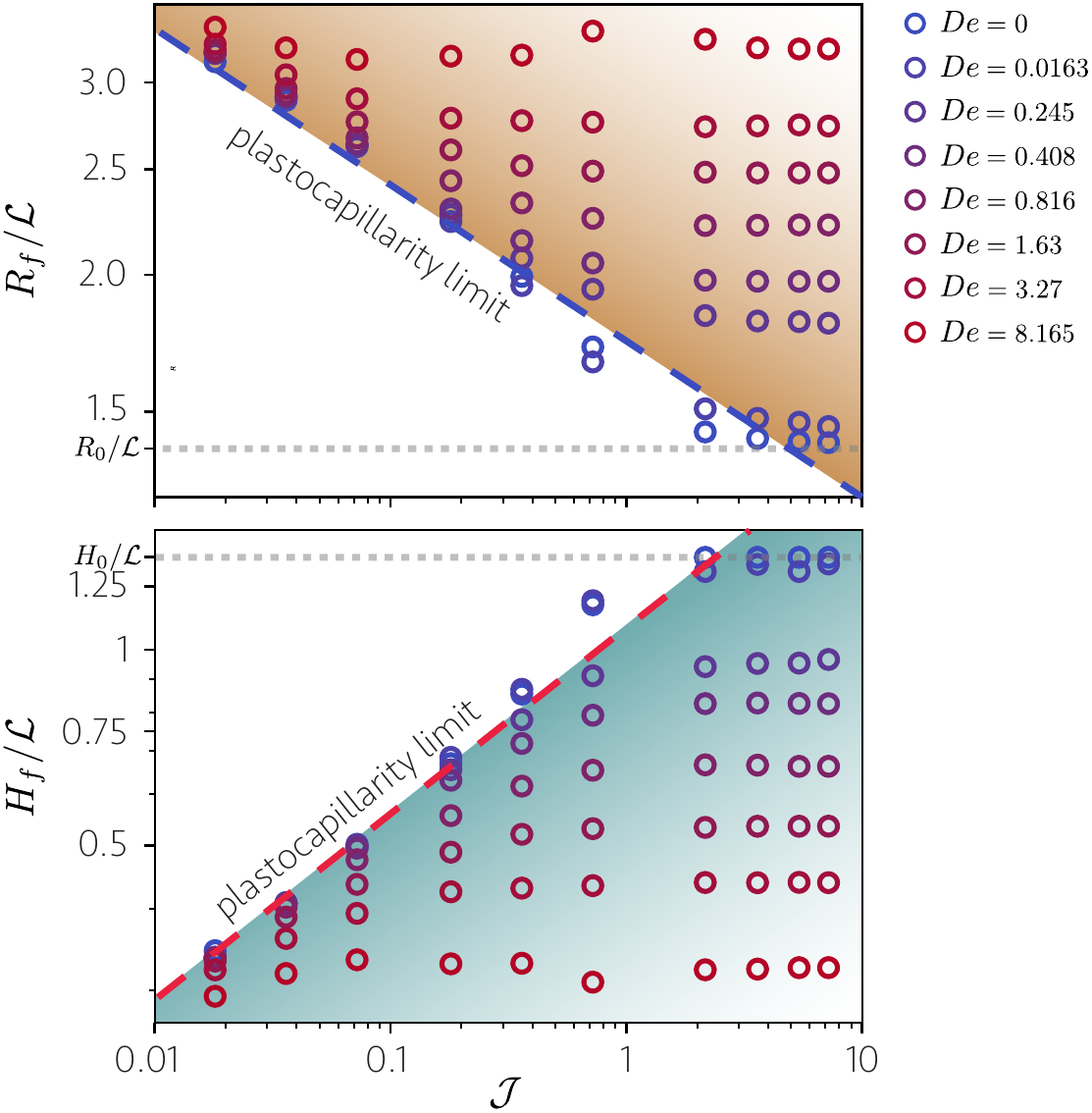}
    \caption{Final radius (top) and height (bottom) of EVP droplets for a range of $\mathcal{J}$ and $De$. The horizontal dashed gray lines show the initial radius and height. The thick blue and red dashed lines correspond to the plastocapillarity limits (see section~\ref{sec:vpscaling}).}
    \label{fig:evp_scalinglaw}
\end{figure}

Finally, we quantify the Deborah number effects on the final radius of the droplet to complete the picture shown in section \ref{sec:vpscaling}. Figure \ref{fig:evp_scalinglaw} shows the final radius of droplet versus $\mathcal{J}$ for different values of $De$. The limit $De = 0$ corresponds to a viscoplastic limit, also shown in figure \ref{fig:vp_valid_mazi_theory} and the dashed lines represent the theoretical scaling for this case. As $De$ increases, the droplet spreads more, particularly for higher values of $\mathcal{J}$. For higher $De$ values, we also see a decrease in the critical value of $\mathcal{J}$ from which the spreading stays almost constant. This happens because, as we saw earlier, the elasticity promotes unyielded droplets, and for droplets that are already unyielded, further increasing the value of $\mathcal{J}$ does not introduce further changes. For the elastocapillary limit, when $De$ and $\mathcal{J}$ are large, balancing the surface tension and elastic forces results in scaling laws for the final radius and height of the droplet. At the stoppage moment, the surface tension force can be estimated as $F_{\sigma} = \sigma \mathcal{H}_f^2 / \mathcal{R}_f$. Balancing this with the elastic forces, estimated as $F_e = (\mu_p / \lambda) \mathcal{R}_f^2$, and a for a given volume of the droplet $\mathcal{V} \sim \mathcal{L}^3 \sim \mathcal{H}_f \mathcal{R}_f^2$, we arrive at $\mathcal{R}_f / \mathcal{L} \sim (Oh_p / De)^{-1/7}$ and $\mathcal{H}_f / \mathcal{L} \sim (Oh_p / De)^{2/7}$. These formulations are similar to those for the plastocapillarity limit, except that the plastocapillary number (yield stress) is replaced by elastocapillary number (elastic module). In appendix~\ref{appC}, we test these scaling laws by looking at the final radius and height as a function of $De$.


\pagebreak
\section{Conclusions}\label{sec:conclusion}
We numerically investigated the spreading of elastoviscoplastic fluids under surface tension forces. Direct numerical simulations are performed using the EVP model of Saramito to understand how different non-dimensional groups influence the spreading. 
We focused our study on the importance of two non-dimensional groups: the plastocapillary number, $\mathcal{J}$, and the Deborah number, $De$. 
We confirm that increasing the plastocapillary number in EVP spreading, reduces the droplet spreading, as previously explained for the viscoplastic limit~\cite{Jalaal2021,van2023viscoplastic}. The influence of elasticity on spreading can also be significant. Increasing the Deborah number (while other parameters are fixed) promotes more spreading. This effect is associated with the polymeric relaxation time, which delays the development of the internal droplet stresses, allowing the droplet to spread with less resistance. 

\begin{figure}[!htb]
	\begin{center}
        \includegraphics[width=0.7\linewidth]{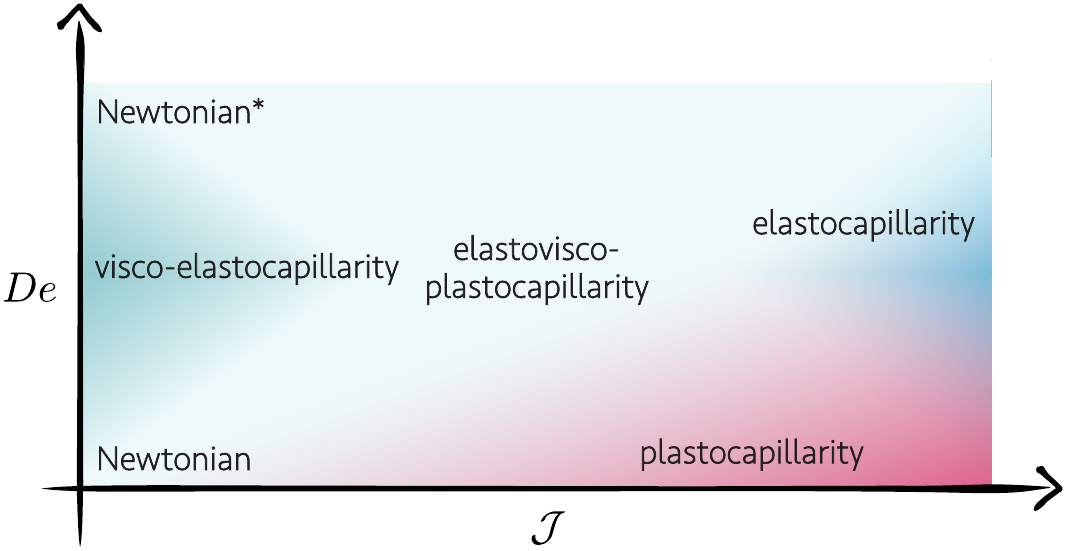}
		\caption{$De-\mathcal{J}$ parameter space and its different limits. The general case of EVP spreading reaches different regimes, depending on which rheological factor (elasticity or plasticity) dominates. Note that the high $De$ regime of Newtonian fluids is for finite $Oh_p$.}
		\label{fig:phasemap}
	\end{center}
\end{figure}

Overall, for fixed (solvent and polymeric) Ohnesorge numbers, the $De-\mathcal{J}$ parameter space covers a range of regimes, including visco-elastocapillarity, elastocapillarity, and plastocapillarity, which can be further studied by the model presented here. Figure~\ref{fig:phasemap} presents a (schematic) overview of these limits.

The spreading of EVP fluids plays a key role in many industrial applications, such as coating and 3D printing. Our work, based on a continuum model, sheds light on how elastic and plastic rheological properties alter viscous spreading. The study can be further extended in many different ways. Firstly, further experimental studies are required. The present experimental data~\cite{Jalaal2021} are limited to viscoplastic limits and are not sufficient to compare with the simulations here. Ideally, experiments in which the values of plastocapillary and the Deborah numbers can be systematically changed should be performed. This, however, could be a difficult task as chemical or physical changes in microstructure often change the yield stress, viscosity, and elastic moduli of the material, simultaneously. 

The present article focuses on the problem of spreading. Still, in principle, the computational framework presented here could be used to study and analyze a range of capillary-driven phenomena such as bubble or drop coalescence~\cite{kern2022viscoplastic,dekker2022elasticity,fardin2022spreading,oratis2023coalescence,rajput2023sub}, pinch-off~\cite{ingremeau2013stretching,rajesh2022pinch}, multi-component systems like drops on liquid-infused surfaces~\cite{sadullah2018drop,kreder2018film} or soft wetting~\cite{style2017elastocapillarity,andreotti2020statics, pandey2020singular,kim2021measuring}, and non-axisymmetric shapes~\cite{friedrich2020corner,}. Finally, the model can be extended for materials with more complicated rheological properties such as thixotropy~\cite{mewis2009thixotropy,Oishi2017, sen2021thixotropy} and also for wetting of dry surfaces, when a triple line exists~\cite{martouzet2021dynamic}.



\section*{Acknowledgements}
The authors would like to thank Daniel Bonn, Antoine Gaillard, Emad Chaparian, and Vatsal Sanjay for insightful discussions.
We also thank the financial support given by Sao Paulo Research Foundation (FAPESP) grants \#2013/07375-0, \#2019/01811-9 and \#2021/14953-6, and the National Council for Scientific and Technological Development (CNPq), grants \#305383/2019-1 and \#304095/2018-4. 

\printbibliography

@article{Viezel2020,
  doi = {10.1016/j.jnnfm.2020.104338},
  url = {https://doi.org/10.1016/j.jnnfm.2020.104338},
  year = {2020},
  month = nov,
  publisher = {Elsevier {BV}},
  volume = {285},
  pages = {104338},
  author = {C. Viezel and M.F. Tom{\'{e}} and F.T. Pinho and S. McKee},
  title = {An Oldroyd-B solver for vanishingly small values of the viscosity ratio: Application to unsteady free surface flows},
  journal = {Journal of Non-Newtonian Fluid Mechanics}
}

@article{LopezHerrera2019,
  doi = {10.1016/j.jnnfm.2018.10.012},
  url = {https://doi.org/10.1016/j.jnnfm.2018.10.012},
  year = {2019},
  month = feb,
  publisher = {Elsevier {BV}},
  volume = {264},
  pages = {144--158},
  author = {J.M. L{\'{o}}pez-Herrera and S. Popinet and A.A. Castrej{\'{o}}n-Pita},
  title = {An adaptive solver for viscoelastic incompressible two-phase problems applied to the study of the splashing of weakly viscoelastic droplets},
  journal = {Journal of Non-Newtonian Fluid Mechanics}
}

@article{Afonso2011,
  doi = {10.1017/jfm.2011.84},
  url = {https://doi.org/10.1017/jfm.2011.84},
  year = {2011},
  month = apr,
  publisher = {Cambridge University Press ({CUP})},
  volume = {677},
  pages = {272--304},
  author = {A. M. Afonso and P. J. Oliveira and F. T. Pinho and M. A. Alves},
  title = {Dynamics of high-Deborah-number entry flows: a numerical study},
  journal = {Journal of Fluid Mechanics}
}

@article{Snoeijer2020,
  doi = {10.1098/rspa.2020.0419},
  url = {https://doi.org/10.1098/rspa.2020.0419},
  year = {2020},
  month = nov,
  publisher = {The Royal Society},
  volume = {476},
  number = {2243},
  author = {J. H. Snoeijer and A. Pandey and M. A. Herrada and J. Eggers},
  title = {The relationship between viscoelasticity and elasticity},
  journal = {Proceedings of the Royal Society A: Mathematical,  Physical and Engineering Sciences}
}

@article{Jalaal2021,
  doi = {10.1017/jfm.2020.886},
  url = {https://doi.org/10.1017/jfm.2020.886},
  year = {2021},
  month = mar,
  publisher = {Cambridge University Press ({CUP})},
  volume = {914},
  author = {Maziyar Jalaal and Boris Stoeber and Neil J. Balmforth},
  title = {Spreading of viscoplastic droplets},
  journal = {Journal of Fluid Mechanics}
}

@article{Saramito2007,
  doi = {10.1016/j.jnnfm.2007.04.004},
  url = {https://doi.org/10.1016/j.jnnfm.2007.04.004},
  year = {2007},
  month = aug,
  publisher = {Elsevier {BV}},
  volume = {145},
  number = {1},
  pages = {1--14},
  author = {Pierre Saramito},
  title = {A new constitutive equation for elastoviscoplastic fluid flows},
  journal = {Journal of Non-Newtonian Fluid Mechanics}
}

@article{Saramito2009,
  doi = {10.1016/j.jnnfm.2008.12.001},
  url = {https://doi.org/10.1016/j.jnnfm.2008.12.001},
  year = {2009},
  month = may,
  publisher = {Elsevier {BV}},
  volume = {158},
  number = {1-3},
  pages = {154--161},
  author = {Pierre Saramito},
  title = {A new elastoviscoplastic model based on the Herschel{\textendash}Bulkley viscoplastic model},
  journal = {Journal of Non-Newtonian Fluid Mechanics}
}

@article{Popinet2015,
  doi = {10.1016/j.jcp.2015.09.009},
  url = {https://doi.org/10.1016/j.jcp.2015.09.009},
  year = {2015},
  month = dec,
  publisher = {Elsevier {BV}},
  volume = {302},
  pages = {336--358},
  author = {St{\'{e}}phane Popinet},
  title = {A quadtree-adaptive multigrid solver for the Serre{\textendash}Green{\textendash}Naghdi equations},
  journal = {Journal of Computational Physics}
}

@article{Izbassarov2020,
  doi = {10.1103/physrevfluids.5.113301},
  url = {https://doi.org/10.1103/physrevfluids.5.113301},
  year = {2020},
  month = nov,
  publisher = {American Physical Society ({APS})},
  volume = {5},
  number = {11},
  author = {D. Izbassarov and O. Tammisola},
  title = {Dynamics of an elastoviscoplastic droplet in a Newtonian medium under shear flow},
  journal = {Physical Review Fluids}
}

@article{Cheddadi2011,
  doi = {10.1140/epje/i2011-11001-4},
  url = {https://doi.org/10.1140/epje/i2011-11001-4},
  year = {2011},
  month = jan,
  publisher = {Springer Science and Business Media {LLC}},
  volume = {34},
  number = {1},
  author = {I. Cheddadi and P. Saramito and B. Dollet and C. Raufaste and F. Graner},
  title = {Understanding and predicting viscous,  elastic,  plastic flows},
  journal = {The European Physical Journal E}
}

@article{Nassar2011,
  doi = {10.1016/j.jnnfm.2011.01.009},
  url = {https://doi.org/10.1016/j.jnnfm.2011.01.009},
  year = {2011},
  month = apr,
  publisher = {Elsevier {BV}},
  volume = {166},
  number = {7-8},
  pages = {386--394},
  author = {Bruno Nassar and Paulo R. de Souza Mendes and M{\^{o}}nica F. Naccache},
  title = {Flow of elasto-viscoplastic liquids through an axisymmetric expansion{\textendash}contraction},
  journal = {Journal of Non-Newtonian Fluid Mechanics}
}

@article{Oishi2017,
  doi = {10.1016/j.jnnfm.2017.07.001},
  url = {https://doi.org/10.1016/j.jnnfm.2017.07.001},
  year = {2017},
  month = sep,
  publisher = {Elsevier {BV}},
  volume = {247},
  pages = {165--177},
  author = {Cassio M. Oishi and Fernando P. Martins and Roney L. Thompson},
  title = {The {\textquotedblleft}avalanche effect{\textquotedblright} of an elasto-viscoplastic thixotropic material on an inclined plane},
  journal = {Journal of Non-Newtonian Fluid Mechanics}
}

@article{Luu2009,
  doi = {10.1017/s0022112009007198},
  url = {https://doi.org/10.1017/s0022112009007198},
  year = {2009},
  month = jul,
  publisher = {Cambridge University Press ({CUP})},
  volume = {632},
  pages = {301--327},
  author = {LI-HUA LUU and YO\"{E}L FORTERRE},
  title = {Drop impact of yield-stress fluids},
  journal = {Journal of Fluid Mechanics}
}

@article{oishi2019normal,
  title={Normal and oblique drop impact of yield stress fluids with thixotropic effects},
  author={Oishi, Cassio M and Thompson, Roney L and Martins, Fernando P},
  journal={Journal of Fluid Mechanics},
  volume={876},
  pages={642--679},
  year={2019},
  publisher={Cambridge University Press}
}

@article{Oishi2019,
  doi = {10.1063/1.5129640},
  url = {https://doi.org/10.1063/1.5129640},
  year = {2019},
  month = dec,
  publisher = {{AIP} Publishing},
  volume = {31},
  number = {12},
  pages = {123109},
  author = {C. M. Oishi and R. L. Thompson and F. P. Martins},
  title = {Impact of capillary drops of complex fluids on a solid surface},
  journal = {Physics of Fluids}
}

@article{Fraggedakis2016,
  doi = {10.1016/j.jnnfm.2016.09.001},
  url = {https://doi.org/10.1016/j.jnnfm.2016.09.001},
  year = {2016},
  month = oct,
  publisher = {Elsevier {BV}},
  volume = {236},
  pages = {104--122},
  author = {D. Fraggedakis and Y. Dimakopoulos and J. Tsamopoulos},
  title = {Yielding the yield stress analysis: A thorough comparison of recently proposed elasto-visco-plastic ({EVP}) fluid models},
  journal = {Journal of Non-Newtonian Fluid Mechanics}
}

@article{Mendes2011,
  doi = {10.1039/c0sm01021a},
  url = {https://doi.org/10.1039/c0sm01021a},
  year = {2011},
  publisher = {Royal Society of Chemistry ({RSC})},
  volume = {7},
  number = {6},
  pages = {2471},
  author = {Paulo R. de Souza Mendes},
  title = {Thixotropic elasto-viscoplastic model for structured fluids},
  journal = {Soft Matter}
}

@article{Barnes1998,
title = {The yield stress-a review or `\textpi \textalpha \textnu \texttau \textalpha \text{ } \textrho \textepsilon \textiota' everything flows?},
journal = {Journal of Non-Newtonian Fluid Mechanics},
volume = {81},
number = {1},
pages = {133-178},
year = {1999},
issn = {0377-0257},
doi = {https://doi.org/10.1016/S0377-0257(98)00094-9},
url = {https://www.sciencedirect.com/science/article/pii/S0377025798000949},
author = {Howard A. Barnes}
}

@article{Derby2010,
  doi = {10.1146/annurev-matsci-070909-104502},
  url = {https://doi.org/10.1146/annurev-matsci-070909-104502},
  year = {2010},
  month = jun,
  publisher = {Annual Reviews},
  volume = {40},
  number = {1},
  pages = {395--414},
  author = {Brian Derby},
  title = {Inkjet Printing of Functional and Structural Materials: Fluid Property Requirements,  Feature Stability,  and Resolution},
  journal = {Annual Review of Materials Research}
}

@article{Thompson2014,
  doi = {10.1017/jfm.2014.621},
  url = {https://doi.org/10.1017/jfm.2014.621},
  year = {2014},
  month = nov,
  publisher = {Cambridge University Press ({CUP})},
  volume = {761},
  pages = {261--281},
  author = {Alice B. Thompson and Carl R. Tipton and Anne Juel and Andrew L. Hazel and Mark Dowling},
  title = {Sequential deposition of overlapping droplets to form a liquid line},
  journal = {Journal of Fluid Mechanics}
}

@article{Mackay2018,
  doi = {10.1122/1.5037687},
  url = {https://doi.org/10.1122/1.5037687},
  year = {2018},
  month = nov,
  publisher = {Society of Rheology},
  volume = {62},
  number = {6},
  pages = {1549--1561},
  author = {Michael E. Mackay},
  title = {The importance of rheological behavior in the additive manufacturing technique material extrusion},
  journal = {Journal of Rheology}
}

@article{Bergemann2018,
  doi = {10.1017/jfm.2018.127},
  url = {https://doi.org/10.1017/jfm.2018.127},
  year = {2018},
  month = mar,
  publisher = {Cambridge University Press ({CUP})},
  volume = {843},
  pages = {1--28},
  author = {Nico Bergemann and Anne Juel and Matthias Heil},
  title = {Viscous drops on a layer of the same fluid: from sinking,  wedging and spreading to their long-time evolution},
  journal = {Journal of Fluid Mechanics}
}

@article{Jalaal2019,
  doi = {10.1017/jfm.2019.734},
  url = {https://doi.org/10.1017/jfm.2019.734},
  year = {2019},
  month = oct,
  publisher = {Cambridge University Press ({CUP})},
  volume = {880},
  pages = {430--440},
  author = {Maziyar Jalaal and Carola Seyfert and Jacco H. Snoeijer},
  title = {Capillary ripples in thin viscous films},
  journal = {Journal of Fluid Mechanics}
}

@article{Papanastasiou1987,
  doi = {10.1122/1.549926},
  url = {https://doi.org/10.1122/1.549926},
  year = {1987},
  month = jul,
  publisher = {Society of Rheology},
  volume = {31},
  number = {5},
  pages = {385--404},
  author = {Tasos C. Papanastasiou},
  title = {Flows of Materials with Yield},
  journal = {Journal of Rheology}
}

@article{Frigaard2005,
  doi = {10.1016/j.jnnfm.2005.01.003},
  url = {https://doi.org/10.1016/j.jnnfm.2005.01.003},
  year = {2005},
  month = apr,
  publisher = {Elsevier {BV}},
  volume = {127},
  number = {1},
  pages = {1--26},
  author = {I.A. Frigaard and C. Nouar},
  title = {On the usage of viscosity regularisation methods for visco-plastic fluid flow computation},
  journal = {Journal of Non-Newtonian Fluid Mechanics}
}

@BOOK{Tryggvason2011-book,
  title     = "Direct numerical simulations of gas-liquid multiphase flows",
  author    = "Tryggvason, Gretar and Scardovelli, Ruben and Zaleski, Stephane",
  publisher = "Cambridge University Press",
  month     =  mar,
  year      =  2011,
  address   = "Cambridge, England"
}

@misc{Popinet2013-Basilisk,   
    title = {Basilisk C},   
    url = {http://basilisk.fr},   
    author = {S. Popinet and Collaborators},   
    year = {2013-2021},   
    note = {Accessed on Month Day, Year} 
}

@article{Hirt1981,
  doi = {10.1016/0021-9991(81)90145-5},
  url = {https://doi.org/10.1016/0021-9991(81)90145-5},
  year = {1981},
  month = jan,
  publisher = {Elsevier {BV}},
  volume = {39},
  number = {1},
  pages = {201--225},
  author = {C.W Hirt and B.D Nichols},
  title = {Volume of fluid ({VOF}) method for the dynamics of free boundaries},
  journal = {Journal of Computational Physics}
}

@article{popinet2003gerris,
  title={Gerris: a tree-based adaptive solver for the incompressible Euler equations in complex geometries},
  author={Popinet, St{\'e}phane},
  journal={Journal of computational physics},
  volume={190},
  number={2},
  pages={572--600},
  year={2003},
  publisher={Elsevier}
}

@article{popinet2009accurate,
  title={An accurate adaptive solver for surface-tension-driven interfacial flows},
  author={Popinet, St{\'e}phane},
  journal={Journal of Computational Physics},
  volume={228},
  number={16},
  pages={5838--5866},
  year={2009},
  publisher={Elsevier}
}

@article{van2023viscoplastic,
  title={Viscoplastic lines: printing a single filament of yield stress material on a surface},
  author={van der Kolk, Jesse and Tieman, Dani{\"e}l and Jalaal, Maziyar},
  journal={Journal of Fluid Mechanics},
  volume={958},
  pages={A34},
  year={2023},
  publisher={Cambridge University Press}
}

@article{friedrich2020corner,
  title={Corner accuracy in direct ink writing with support material},
  author={Friedrich, Leanne and Begley, Matthew},
  journal={Bioprinting},
  volume={19},
  pages={e00086},
  year={2020},
  publisher={Elsevier}
}

@article{milazzo20233d,
  title={3D Printability of Silk/Hydroxyapatite Composites for Microprosthetic Applications},
  author={Milazzo, Mario and Fitzpatrick, Vincent and Owens, Crystal E and Carraretto, Igor M and McKinley, Gareth H and Kaplan, David L and Buehler, Markus J},
  journal={ACS Biomaterials Science \& Engineering},
  volume={9},
  number={3},
  pages={1285--1295},
  year={2023},
  publisher={ACS Publications}
}

@article{rauzan2018particle,
  title={Particle-free emulsions for 3D printing elastomers},
  author={Rauzan, Brittany M and Nelson, Arif Z and Lehman, Sean E and Ewoldt, Randy H and Nuzzo, Ralph G},
  journal={Advanced Functional Materials},
  volume={28},
  number={21},
  pages={1707032},
  year={2018},
  publisher={Wiley Online Library}
}

@article{saadi2022direct,
  title={Direct ink writing: a 3D printing technology for diverse materials},
  author={Saadi, MASR and Maguire, Alianna and Pottackal, Neethu T and Thakur, Md Shajedul Hoque and Ikram, Maruf Md and Hart, A John and Ajayan, Pulickel M and Rahman, Muhammad M},
  journal={Advanced Materials},
  volume={34},
  number={28},
  pages={2108855},
  year={2022},
  publisher={Wiley Online Library}
}

@article{garcia2023fourier,
  title={Fourier-transform rheology and printability maps of complex fluids for three-dimensional printing},
  author={Garc{\'\i}a-Tu{\~n}{\'o}n, Esther and Agrawal, Rishav and Ling, Bin and Dennis, David JC},
  journal={Physics of Fluids},
  volume={35},
  number={1},
  pages={017113},
  year={2023},
  publisher={AIP Publishing LLC}
}

@article{tanner1979spreading,
  title={The spreading of silicone oil drops on horizontal surfaces},
  author={Tanner, LH},
  journal={Journal of Physics D: Applied Physics},
  volume={12},
  number={9},
  pages={1473},
  year={1979},
  publisher={IOP Publishing}
}

@article{lohse2022fundamental,
  title={Fundamental fluid dynamics challenges in inkjet printing},
  author={Lohse, Detlef},
  journal={Annual Review of Fluid Mechanics},
  volume={54},
  year={2022},
  publisher={Annual Reviews}
}

@article{ewoldt2021designing,
  title={Designing complex fluids},
  author={Ewoldt, Randy H and Saengow, Chaimongkol},
  journal={Annual Review of Fluid Mechanics},
  volume={54},
  year={2021},
  publisher={Annual Reviews}
}

@article{saramito2017progress,
  title={Progress in numerical simulation of yield stress fluid flows},
  author={Saramito, Pierre and Wachs, Anthony},
  journal={Rheologica Acta},
  volume={56},
  number={3},
  pages={211--230},
  year={2017},
  publisher={Springer}
}

@article{dimitriou2014comprehensive,
  title={A comprehensive constitutive law for waxy crude oil: a thixotropic yield stress fluid},
  author={Dimitriou, Christopher J and McKinley, Gareth H},
  journal={Soft Matter},
  volume={10},
  number={35},
  pages={6619--6644},
  year={2014},
  publisher={Royal Society of Chemistry}
}

@article{bonn2009wetting,
  title={Wetting and spreading},
  author={Bonn, Daniel and Eggers, Jens and Indekeu, Joseph and Meunier, Jacques and Rolley, Etienne},
  journal={Reviews of modern physics},
  volume={81},
  number={2},
  pages={739},
  year={2009},
  publisher={APS}
}

@article{gorin2022universal,
  title={Universal aspects of droplet spreading dynamics in Newtonian and non-Newtonian fluids},
  author={Gorin, Benjamin and Di Mauro, Gabrielle and Bonn, Daniel and Kellay, Hamid},
  journal={Langmuir},
  volume={38},
  number={8},
  pages={2608--2613},
  year={2022},
  publisher={ACS Publications}
}

@article{bergeron2000controlling,
  title={Controlling droplet deposition with polymer additives},
  author={Bergeron, Vance and Bonn, Daniel and Martin, Jean Yves and Vovelle, Louis},
  journal={Nature},
  volume={405},
  number={6788},
  pages={772--775},
  year={2000},
  publisher={Nature Publishing Group UK London}
}

@article{wang2015dynamic,
  title={Dynamic wetting of viscoelastic droplets},
  author={Wang, Yuli and Minh, Do-Quang and Amberg, Gustav},
  journal={Physical Review E},
  volume={92},
  number={4},
  pages={043002},
  year={2015},
  publisher={APS}
}

@article{wang2017impact,
  title={Impact of viscoelastic droplets},
  author={Wang, Yuli and Do-Quang, Minh and Amberg, Gustav},
  journal={Journal of Non-Newtonian Fluid Mechanics},
  volume={243},
  pages={38--46},
  year={2017},
  publisher={Elsevier}
}

@article{izbassarov2016effects,
  title={Effects of viscoelasticity on drop impact and spreading on a solid surface},
  author={Izbassarov, Daulet and Muradoglu, Metin},
  journal={Physical Review Fluids},
  volume={1},
  number={2},
  pages={023302},
  year={2016},
  publisher={APS}
}

@article{saidi2010influence,
  title={Influence of yield stress on the fluid droplet impact control},
  author={Sa{\"\i}di, Alireza and Martin, C{\'e}line and Magnin, Albert},
  journal={Journal of Non-Newtonian Fluid Mechanics},
  volume={165},
  number={11-12},
  pages={596--606},
  year={2010},
  publisher={Elsevier}
}

@article{blackwell2015sticking,
  title={Sticking and splashing in yield-stress fluid drop impacts on coated surfaces},
  author={Blackwell, Brendan C and Deetjen, Marc E and Gaudio, Joseph E and Ewoldt, Randy H},
  journal={Physics of Fluids},
  volume={27},
  number={4},
  pages={043101},
  year={2015},
  publisher={AIP Publishing LLC}
}

@article{saidi2011effects,
  title={Effects of surface properties on the impact process of a yield stress fluid drop},
  author={Sa{\"\i}di, Alireza and Martin, C{\'e}line and Magnin, Albert},
  journal={Experiments in fluids},
  volume={51},
  number={1},
  pages={211--224},
  year={2011},
  publisher={Springer}
}

@article{martouzet2021dynamic,
  title={Dynamic arrest during the spreading of a yield stress fluid drop},
  author={Martouzet, Gr{\'e}goire and J{\o}rgensen, Loren and Pelet, Yoann and Biance, Anne-Laure and Barentin, Catherine},
  journal={Physical Review Fluids},
  volume={6},
  number={4},
  pages={044006},
  year={2021},
  publisher={APS}
}

@article{d2022spreading,
  title={Spreading of droplets under various gravitational accelerations},
  author={D’Angelo, Olfa and Kuthe, Felix and van Nieuwland, Kasper and Ederveen Janssen, Clint and Voigtmann, Thomas and Jalaal, Maziyar},
  journal={Review of Scientific Instruments},
  volume={93},
  number={11},
  pages={115103},
  year={2022},
  publisher={AIP Publishing LLC}
}

@article{izbassarov2021effect,
  title={Effect of finite Weissenberg number on turbulent channel flows of an elastoviscoplastic fluid},
  author={Izbassarov, Daulet and Rosti, Marco E and Brandt, Luca and Tammisola, Outi},
  journal={Journal of Fluid Mechanics},
  volume={927},
  pages={A45},
  year={2021},
  publisher={Cambridge University Press}
}

@article{chaparian2020yield,
  title={Yield-stress fluids in porous media: a comparison of viscoplastic and elastoviscoplastic flows},
  author={Chaparian, Emad and Izbassarov, Daulet and De Vita, Francesco and Brandt, Luca and Tammisola, Outi},
  journal={Meccanica},
  volume={55},
  pages={331--342},
  year={2020},
  publisher={Springer}
}

@article{chaparian2020particle,
  title={Particle migration in channel flow of an elastoviscoplastic fluid},
  author={Chaparian, Emad and Ardekani, Mehdi N and Brandt, Luca and Tammisola, Outi},
  journal={Journal of Non-Newtonian Fluid Mechanics},
  volume={284},
  pages={104376},
  year={2020},
  publisher={Elsevier}
}

@article{de2019oscillations,
  title={Oscillations of small bubbles and medium yielding in elastoviscoplastic fluids},
  author={De Corato, Marco and Saint-Michel, Brice and Makrigiorgos, George and Dimakopoulos, Yannis and Tsamopoulos, John and Garbin, Valeria},
  journal={Physical Review Fluids},
  volume={4},
  number={7},
  pages={073301},
  year={2019},
  publisher={APS}
}

@article{moschopoulos2021concept,
  title={The concept of elasto-visco-plasticity and its application to a bubble rising in yield stress fluids},
  author={Moschopoulos, Pantelis and Spyridakis, Alexandros and Varchanis, Stylianos and Dimakopoulos, Yannis and Tsamopoulos, John},
  journal={Journal of Non-Newtonian Fluid Mechanics},
  volume={297},
  pages={104670},
  year={2021},
  publisher={Elsevier}
}

@article{kordalis2021investigation,
  title={Investigation of the extensional properties of elasto-visco-plastic materials in cross-slot geometries},
  author={Kordalis, A and Varchanis, S and Ioannou, G and Dimakopoulos, Y and Tsamopoulos, J},
  journal={Journal of Non-Newtonian Fluid Mechanics},
  volume={296},
  pages={104627},
  year={2021},
  publisher={Elsevier}
}

@article{fraggedakis2016yielding,
  title={Yielding the yield-stress analysis: a study focused on the effects of elasticity on the settling of a single spherical particle in simple yield-stress fluids},
  author={Fraggedakis, D and Dimakopoulos, Y and Tsamopoulos, J},
  journal={Soft matter},
  volume={12},
  number={24},
  pages={5378--5401},
  year={2016},
  publisher={Royal Society of Chemistry}
}

@article{de2018elastoviscoplastic,
  title={Elastoviscoplastic flows in porous media},
  author={Duffo, L and Hormozi, S},
  journal={Journal of Non-Newtonian Fluid Mechanics},
  volume={258},
  pages={10--21},
  year={2018},
  publisher={Elsevier}
}

@article{kern2022viscoplastic,
  title={Viscoplastic sessile drop coalescence},
  author={Kern, Vanessa R and S{\ae}ter, Torstein and Carlson, Andreas},
  journal={Physical Review Fluids},
  volume={7},
  number={8},
  pages={L081601},
  year={2022},
  publisher={APS}
}

@article{sen2021thixotropy,
  title={Thixotropy in viscoplastic drop impact on thin films},
  author={Sen, Samya and Morales, Anthony G and Ewoldt, Randy H},
  journal={Physical Review Fluids},
  volume={6},
  number={4},
  pages={043301},
  year={2021},
  publisher={APS}
}

@article{sen2020,
  title={Viscoplastic drop impact on thin films},
  author={Sen, Samya and Morales, Anthony G and Ewoldt, Randy H},
  journal={Journal of Fluid Mechanics},
  volume={891},
  pages={A27},
  year={2020},
  publisher={Cambridge University Press}
}

@article{sanjay2021bursting,
  title={Bursting bubble in a viscoplastic medium},
  author={Sanjay, Vatsal and Lohse, Detlef and Jalaal, Maziyar},
  journal={Journal of fluid mechanics},
  volume={922},
  pages={A2},
  year={2021},
  publisher={Cambridge University Press}
}

@article{jalaal2023plastocapillarity,
  title={Plastocapillarity},
  author={Jalaal, Maziyar},
  journal={Science Talks},
  volume={6},
  year={2023},
  publisher={Elsevier}
}

@article{frigaard2019simple,
  title={Simple yield stress fluids},
  author={Frigaard, Ian},
  journal={Current Opinion in Colloid \& Interface Science},
  volume={43},
  pages={80--93},
  year={2019},
  publisher={Elsevier}
}

@book{bingham1917investigation,
  title={An investigation of the laws of plastic flow},
  author={Bingham, Eugene Cook},
  number={278},
  year={1917},
  publisher={US Government Printing Office}
}

@article{oldroyd1950formulation,
  title={On the formulation of rheological equations of state},
  author={Oldroyd, James G},
  journal={Proceedings of the Royal Society of London. Series A. Mathematical and Physical Sciences},
  volume={200},
  number={1063},
  pages={523--541},
  year={1950},
  publisher={The Royal Society London}
}

@article{stone2023note,
  title={A note about convected time derivatives for flows of complex fluids},
  author={Stone, Howard A and Shelley, Michael J and Boyko, Evgeniy},
  journal={arXiv preprint arXiv:2304.06449},
  year={2023}
}

@article{hinch2021oldroyd,
  title={Oldroyd B, and not A?},
  author={Hinch, John and Harlen, Oliver},
  journal={Journal of Non-Newtonian Fluid Mechanics},
  volume={298},
  pages={104668},
  year={2021},
  publisher={Elsevier}
}

@article{morozov2015introduction,
  title={Introduction to complex fluids},
  author={Morozov, Alexander and Spagnolie, Saverio E},
  journal={Complex Fluids in Biological Systems: Experiment, Theory, and Computation},
  pages={3--52},
  year={2015},
  publisher={Springer}
}

@article{balmforth2001geophysical,
  title={Geophysical aspects of non-Newtonian fluid mechanics},
  author={Balmforth, NJ and Craster, RV},
  journal={Geomorphological Fluid Mechanics},
  pages={34--51},
  year={2001},
  publisher={Springer}
}

@article{ancey2007plasticity,
  title={Plasticity and geophysical flows: a review},
  author={Ancey, Christophe},
  journal={Journal of non-Newtonian fluid mechanics},
  volume={142},
  number={1-3},
  pages={4--35},
  year={2007},
  publisher={Elsevier}
}

@article{deblais2018pearling,
  title={Pearling Instabilities of a Viscoelastic Thread},
  author={Deblais, A and Velikov, KP and Bonn, D and others},
  journal={Physical Review Letters},
  volume={120},
  number={194501},
  year={2018}
}

@article{clasen2006beads,
  title={The beads-on-string structure of viscoelastic threads},
  author={Clasen, Christian and Eggers, Jens and Fontelos, Marco A and Li, JIE and McKinley, Gareth H},
  journal={Journal of Fluid Mechanics},
  volume={556},
  pages={283--308},
  year={2006},
  publisher={Cambridge University Press}
}

@article{ardekani2010dynamics,
  title={Dynamics of bead formation, filament thinning and breakup in weakly viscoelastic jets},
  author={Ardekani, AM and Sharma, V and McKinley, GH},
  journal={Journal of Fluid Mechanics},
  volume={665},
  pages={46--56},
  year={2010},
  publisher={Cambridge University Press}
}

@article{bhat2010formation,
  title={Formation of beads-on-a-string structures during break-up of viscoelastic filaments},
  author={Bhat, Pradeep P and Appathurai, Santosh and Harris, Michael T and Pasquali, Matteo and McKinley, Gareth H and Basaran, Osman A},
  journal={Nature Physics},
  volume={6},
  number={8},
  pages={625--631},
  year={2010},
  publisher={Nature Publishing Group UK London}
}

@article{mckinley2005visco,
  title={Visco-elasto-capillary thinning and break-up of complex fluids},
  author={McKinley, Gareth H},
  year={2005}
}

@book{kelvin1865elasticity,
  title={On the elasticity and viscosity of metals},
  author={Kelvin, William Thomson Baron},
  year={1865}
}

@article{voigt1892ueber,
  title={Ueber innere Reibung fester K{\"o}rper, insbesondere der Metalle},
  author={Voigt, Woldemar},
  journal={Annalen der Physik},
  volume={283},
  number={12},
  pages={671--693},
  year={1892},
  publisher={Wiley Online Library}
}

@book{lakes1998viscoelastic,
  title={Viscoelastic solids},
  author={Lakes, Roderic S},
  volume={9},
  year={1998},
  publisher={CRC press}
}

@article{bico2018elastocapillarity,
  title={Elastocapillarity: when surface tension deforms elastic solids},
  author={Bico, Jos{\'e} and Reyssat, {\'E}tienne and Roman, Beno{\^\i}t},
  journal={Annual Review of Fluid Mechanics},
  volume={50},
  pages={629--659},
  year={2018},
  publisher={Annual Reviews}
}

@article{style2017elastocapillarity,
  title={Elastocapillarity: Surface tension and the mechanics of soft solids},
  author={Style, Robert W and Jagota, Anand and Hui, Chung-Yuen and Dufresne, Eric R},
  journal={Annual Review of Condensed Matter Physics},
  volume={8},
  pages={99--118},
  year={2017},
  publisher={Annual Reviews}
}

@article{andreotti2020statics,
  title={Statics and dynamics of soft wetting},
  author={Andreotti, Bruno and Snoeijer, Jacco H},
  journal={Annual review of fluid mechanics},
  volume={52},
  pages={285--308},
  year={2020},
  publisher={Annual Reviews}
}

@article{oratis2023coalescence,
  title={Coalescence of bubbles in a viscoelastic liquid},
  author={Oratis, Alexandros T and Bertin, Vincent and Snoeijer, Jacco H},
  journal={arXiv preprint arXiv:2305.01363},
  year={2023}
}

@article{dekker2022elasticity,
  title={When elasticity affects drop coalescence},
  author={Dekker, Pim J and Hack, Michiel A and Tewes, Walter and Datt, Charu and Bouillant, Ambre and Snoeijer, Jacco H},
  journal={Physical review letters},
  volume={128},
  number={2},
  pages={028004},
  year={2022},
  publisher={APS}
}

@article{rajput2023sub,
  title={Sub-Newtonian coalescence in polymeric fluids},
  author={Rajput, Abhineet Singh and Varma, Sarath Chandra and Kumar, Aloke},
  journal={Soft Matter},
  year={2023},
  publisher={Royal Society of Chemistry}
}

@article{mewis2009thixotropy,
  title={Thixotropy},
  author={Mewis, Jan and Wagner, Norman J},
  journal={Advances in colloid and interface science},
  volume={147},
  pages={214--227},
  year={2009},
  publisher={Elsevier}
}

@article{ingremeau2013stretching,
  title={Stretching polymers in droplet-pinch-off experiments},
  author={Ingremeau, Fran{\c{c}}ois and Kellay, Hamid},
  journal={Physical Review X},
  volume={3},
  number={4},
  pages={041002},
  year={2013},
  publisher={APS}
}

@article{rajesh2022pinch,
  title={Pinch-off of bubbles in a polymer solution},
  author={Rajesh, Sreeram and Peddada, Sumukh S and Thi{\'e}venaz, Virgile and Sauret, Alban},
  journal={Journal of Non-Newtonian Fluid Mechanics},
  volume={310},
  pages={104921},
  year={2022},
  publisher={Elsevier}
}

@article{kreder2018film,
  title={Film dynamics and lubricant depletion by droplets moving on lubricated surfaces},
  author={Kreder, Michael J and Daniel, Dan and Tetreault, Adam and Cao, Zhenle and Lemaire, Baptiste and Timonen, Jaakko VI and Aizenberg, Joanna},
  journal={Physical Review X},
  volume={8},
  number={3},
  pages={031053},
  year={2018},
  publisher={APS}
}

@article{sadullah2018drop,
  title={Drop dynamics on liquid-infused surfaces: The role of the lubricant ridge},
  author={Sadullah, Muhammad S and Semprebon, Ciro and Kusumaatmaja, Halim},
  journal={Langmuir},
  volume={34},
  number={27},
  pages={8112--8118},
  year={2018},
  publisher={ACS Publications}
}

@article{kim2021measuring,
  title={Measuring surface tensions of soft solids with huge contact-angle hysteresis},
  author={Kim, Jin Young and Heyden, Stefanie and Gerber, Dominic and Bain, Nicolas and Dufresne, Eric R and Style, Robert W},
  journal={Physical Review X},
  volume={11},
  number={3},
  pages={031004},
  year={2021},
  publisher={APS}
}

@article{pandey2020singular,
  title={Singular nature of the elastocapillary ridge},
  author={Pandey, A and Andreotti, B and Karpitschka, S and Van Zwieten, GJ and Van Brummelen, EH and Snoeijer, JH},
  journal={Physical review X},
  volume={10},
  number={3},
  pages={031067},
  year={2020},
  publisher={APS}
}

@article{fardin2022spreading,
  title={Spreading, pinching, and coalescence: the Ohnesorge units},
  author={Fardin, Marc A and Hautefeuille, Mathieu and Sharma, Vivek},
  journal={Soft Matter},
  volume={18},
  number={17},
  pages={3291--3303},
  year={2022},
  publisher={Royal Society of Chemistry}
}


\appendix

\section*{Appendix: Numerical regularization and grid dependency}

\subsection{Viscoplasticity and viscoelasticity}
\label{appA}
\begin{figure}[h]
    \centering
\includegraphics[width=0.3\textwidth]{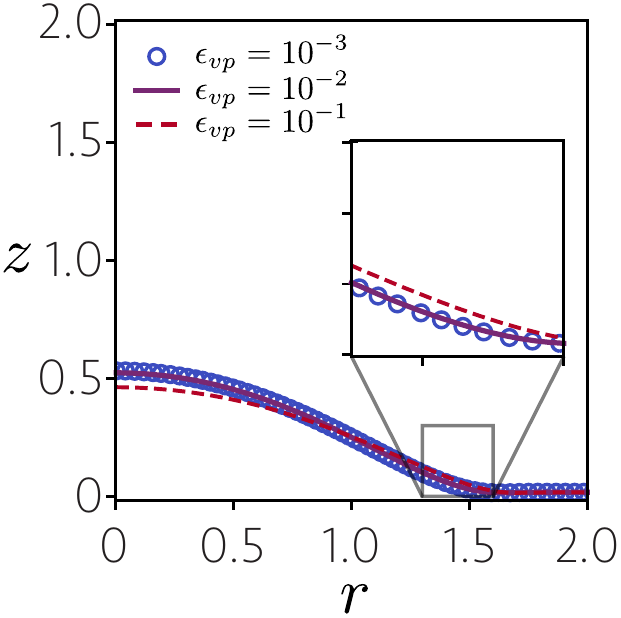}
    \caption{Convergence tests for the Bingham regularization parameter. The plastocapillary number is $\mathcal{J} = 0.18$ and
    the mesh is fixed at Level = 10. 
    }
    \label{fig:vp_valid_reg_mesh2}
\end{figure}

We use a generalized Newtonian Bingham model when $De = 0$, i.e, a viscoplastic material. In this case, the apparent viscosity is replaced by a regularized Bingham-Papanastasiou viscosity \cite{Papanastasiou1987}
\begin{equation}
    \mu_{app} = Oh_s + Oh_p + \frac{\mathcal{J}}{2\norm{\textbf{D}}}\left( 1 - e^{-\norm{\textbf{D}}/\epsilon_{vp}}\right),
\label{eq:vp_regularized}
\end{equation}
where $\epsilon_{vp}$ is the regularization parameter (see~\cite{Frigaard2005}).
Figure \ref{fig:vp_valid_reg_mesh2} shows an example of convergence tests for the viscoplastic regularization parameter 
when $\mathcal{J} = 0.18$, $Oh_s + Oh_p = 0.1$, $De = 0$, $Bo = 0$ and the interface is captured at time $t = 10$. Performing a series of theses tests, we note that all solutions have converged for values $\epsilon_{vp} \leq 10^{-3}$. Therefore, in all viscoplastic simulations reported in this article, we fixed $\epsilon_{vp} =  10^{-3}$.

\begin{figure}[h]
    \centering
\includegraphics[width=0.8\textwidth]{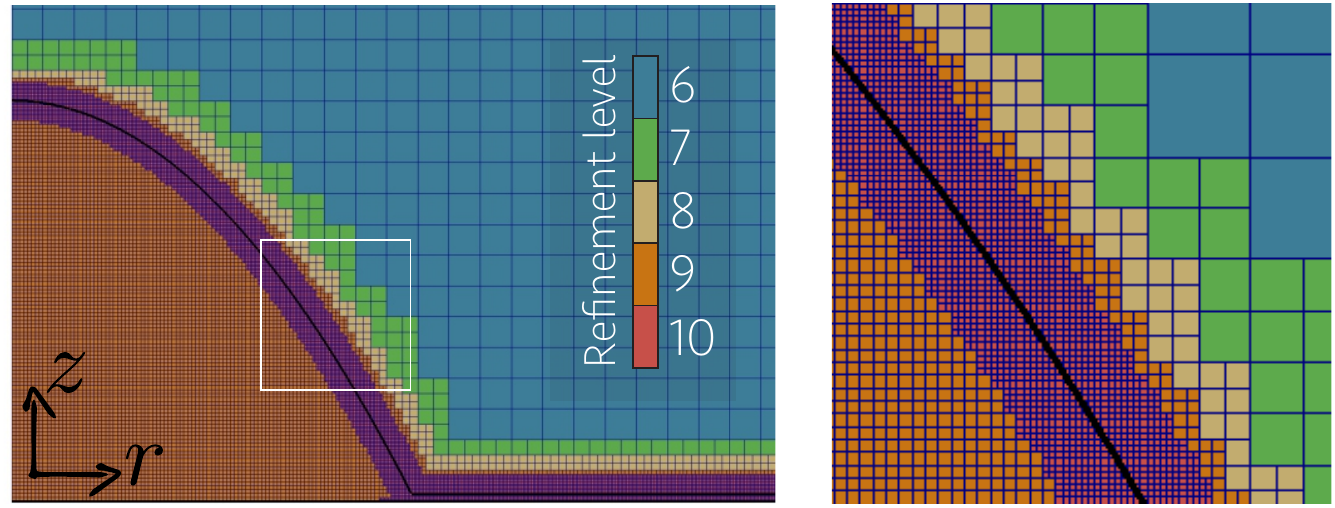}
    \caption{Non-uniform grid generated for the spreading simulation. The right panel shows the magnified view of the white box in the left. Cells are colored by their level of refinement (the computational domain is a square and has a level 0 refinement; see~\cite{popinet2003gerris}). We note that this figure does not capture the entire domain, which continues both in the $r$ and $z$ directions.}
    \label{fig:mesh}
\end{figure}

\begin{figure}[h]
    \centering
\includegraphics[width=1\textwidth]{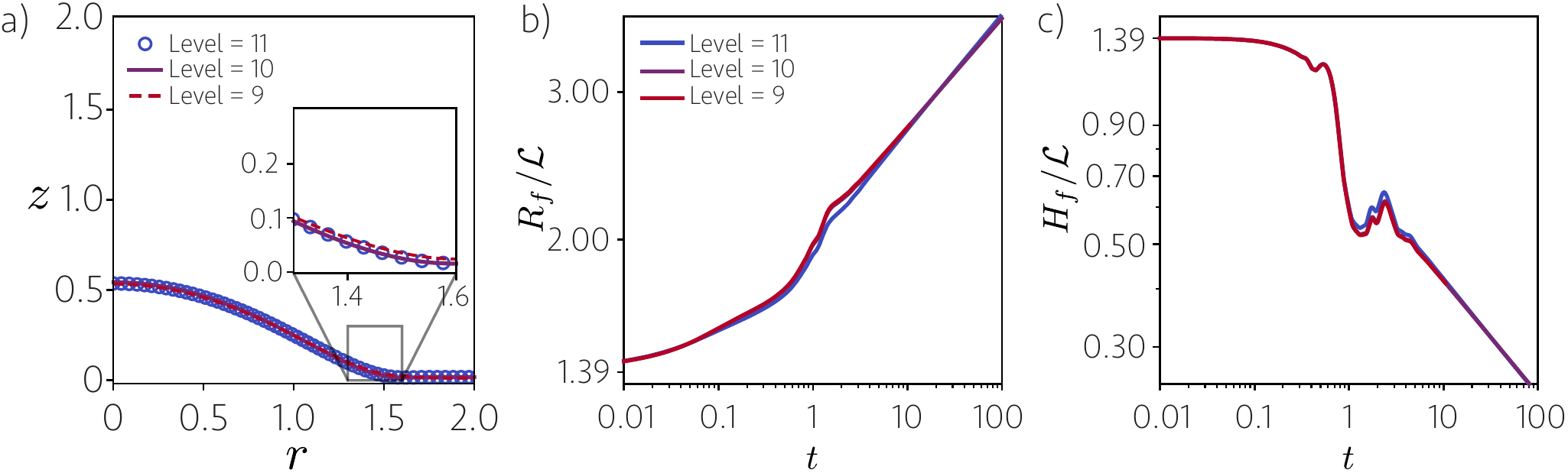}
    \caption{Mesh convergence tests for a Bingham material and a viscoelastic fluid. a) Snapshot of a Bingham droplet at time $t=10$ with fixed $\epsilon_{vp} = 10^{-3}$ and $\mathcal{J} - 0.18$. b) and c) Radius and height evolution for a viscoelastic droplet with $De = 0.41$ and different mesh refinements.}
    \label{fig:vp_valid_reg_mesh}
\end{figure}

An adaptive quadtree mesh is used in all the simulations in this work. An example of non-uniform grids is shown in figure~\ref{fig:mesh}. The maximum level of refinement is initially applied only near the interface, while one level coarser is used everywhere inside the droplet. Over time, the adaptive mesh refinement may also impose the maximum level in regions inside the droplet, based on error estimations on the velocity and volume fractions. In the outer (air) phase, the size of the grids gradually increases from the interface.
We checked for the effect of grid size on the numerical results presented here to ensure the reported outcomes on dynamics of spreading and the droplet shapes are independent of the grid size.
 In Figure \ref{fig:vp_valid_reg_mesh}a, the maximum mesh refinement level is varied for a fixed $\epsilon_{vp} = 10^{-3}$. The change in the interface dynamics is negligible when 
 $\text{Level} \geq 10$. With these results in mind,  all the viscoplastic simulations in this paper are performed with $\epsilon_{vp} = 10^{-3}$ and $\text{Level} = 10$.

We now perform a convergence test for the case of viscoelastic droplets ($\mathcal{J} = 0$ and $De \neq 0$). In this situation, no viscoplastic regularization needs to be used, so we only check for mesh convergence. Figures \ref{fig:vp_valid_reg_mesh}b and \ref{fig:vp_valid_reg_mesh}c shows the radius and height of the droplet over time for different mesh levels with fixed $De = 0.41$. We note that little difference is observed with mesh levels above 9. For this reason, all the viscoelastic simulations performed in this work will use the mesh level 10.

\subsection{Elastoviscoplasticity}
\label{appB}
When $De \neq 0$ and $\mathcal{J} \neq 0$, the full Saramito constitutive equations \eqref{eqAdim3} are used. The software Basilisk already contains a well-tested implementation of the Oldroyd-B model for solving viscoelastic flows given the parameters $\lambda$ (relaxation time) and $\mu_p$ (polymeric viscosity) \cite{LopezHerrera2019}. To make use of this existing module, we re-wrote the Saramito constitutive equation \eqref{eqAdim3} as 
\begin{equation}
    \frac{De}{\alpha} \stackrel{\raisebox{.3ex}{ \hskip -.3cm$\scriptscriptstyle{\bigtriangledown}$}}{\bm{\tau}^p} + \bm{\tau}^p = 2\ \frac{Oh_p}{\alpha}\ \textbf{D}, \hspace{20pt} \text{where} \hspace{20pt} \alpha = \max{\left(\epsilon_{evp}, 1 - \frac{\mathcal{J}}{\norm{\bm{\tau}^p}} \right)},
\label{eq:basilisk_evp}
\end{equation}
and $\epsilon_{evp}$ is a small threshold parameter. We note that equation \eqref{eq:basilisk_evp} has the same form as the traditional Oldroyd-B equation but with non-constant relaxation time and polymeric viscosity coefficients. Therefore, we use the standard Oldroyd-B solver in Basilisk by dynamically setting these two coefficients according to the expressions in equation \eqref{eq:basilisk_evp}.

\begin{figure}[h]
    \centering
\includegraphics[width=0.7\textwidth]{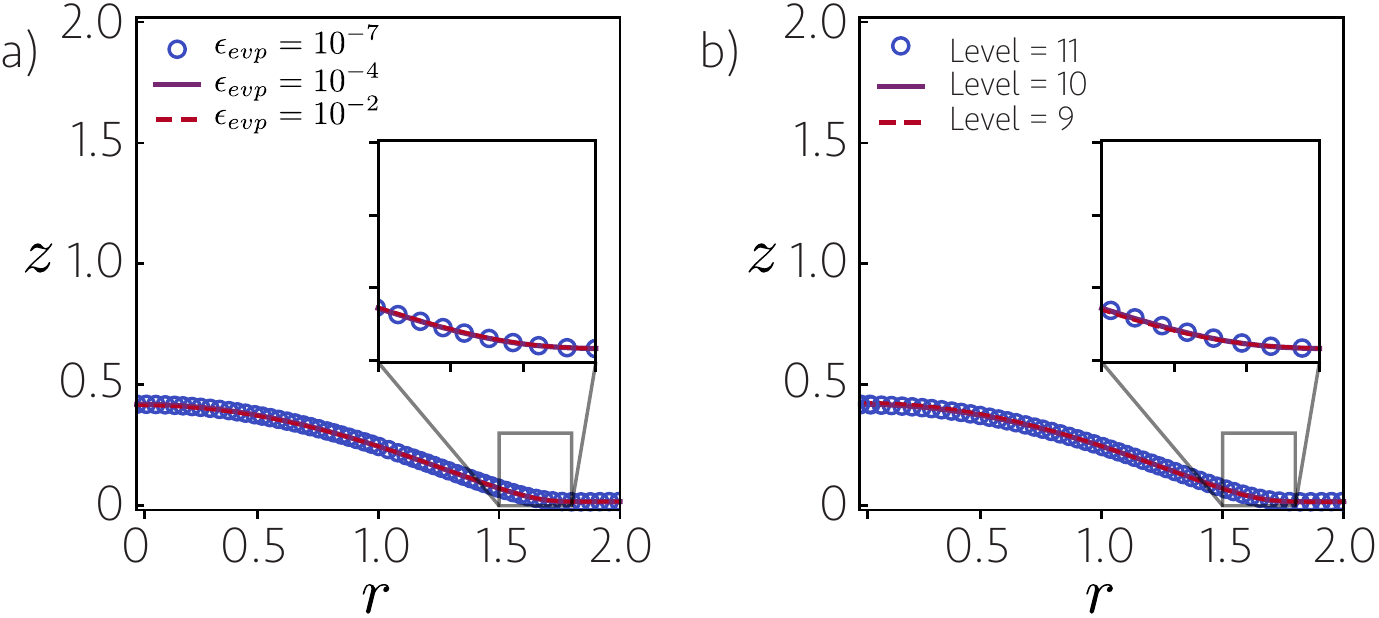}
    \caption{Convergence tests for the EVP regularization parameter (left) and for the mesh refinement parameter (right). In the regularization tests, we keep fixed the mesh level as 10 and in the mesh tests we fix the regularization $\epsilon_{evp} = 10^{-7}$. All simulations are performed with $De = 0.816$ and $\mathcal{J} = 0.18$.
    }
    \label{fig:evp_valid_reg_mesh}
\end{figure}

Figure \ref{fig:evp_valid_reg_mesh} shows convergence tests for the EVP threshold parameter and for the mesh refinement parameter in a simulation with $De = 0.816$ and $\mathcal{J} = 0.18$. We observe that $\epsilon_{evp}$ has negligible influence on the shape of the droplet, even at relatively large values. Regarding mesh convergence, we see that for refinement levels above 9, the results are already very similar. For all the following EVP simulations in this article, we used the combination $\epsilon_{evp} = 10^{-7}$ and Level = 10.


\subsection{Elastocapillarity}
\label{appC}

\begin{figure}[h]
    \centering
\includegraphics[width=0.7\textwidth]{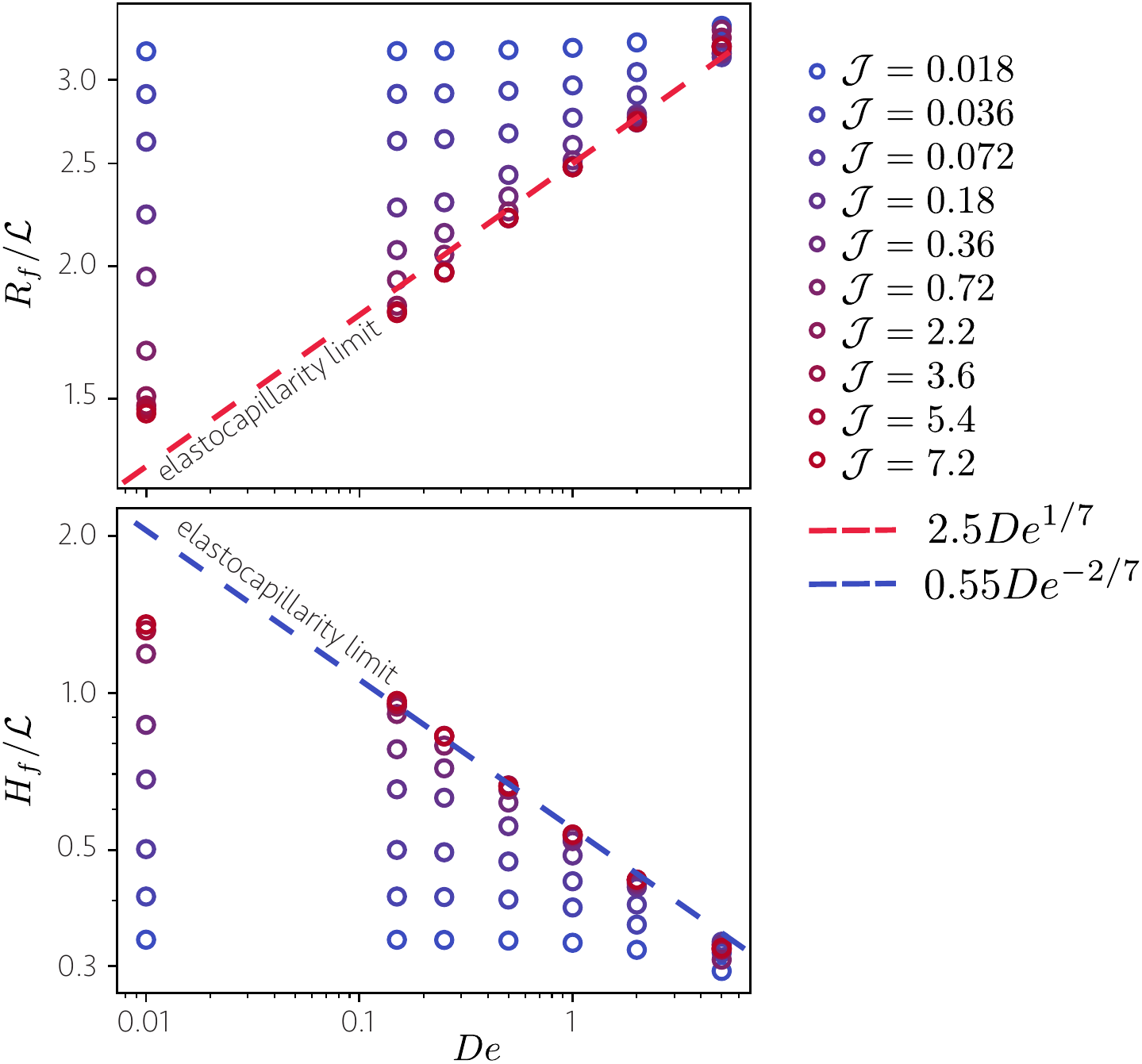}
    \caption{Final radius and height as a function of $De$. The dashed lines show the scaling laws for the elastocapillarity limit, where both $\mathcal{J}$ and $De$ are high enough such that the droplet spreads and stops like a soft solid.
    }
    \label{fig:elastocap}
\end{figure}

Figure~\ref{fig:elastocap} presents the data in Figure~\ref{fig:evp_scalinglaw} as a function of $De$. Note that the pre-factors presented here are simply good fit through the data and are not from asymptotic analysis. A more rigorous analysis like those presented for viscoplastic limits~\cite{Jalaal2021} is needed to find the exact pre-factors. 

\end{document}